\documentclass[journal,twoside]{IEEEtran}

\IEEEoverridecommandlockouts{}
\pdfoutput=1 
\usepackage[utf8]{inputenc} 
\usepackage[T1]{fontenc} 
\usepackage{amsmath}
\usepackage{amsfonts}
\usepackage{amsbsy}
\usepackage{amssymb}
\usepackage{bbm}
\usepackage[italicdiff]{physics}
\usepackage{proba}
\usepackage{enumerate}
\usepackage{cite}
\usepackage{wrapfig}
\usepackage{siunitx}
\usepackage{soul} 
\usepackage{times}
\usepackage{adjustbox}
\usepackage{xfrac}
\usepackage{url}
\usepackage{graphicx}
\usepackage{rotating}
\usepackage{mdframed}
\usepackage{cuted}
\usepackage{xcolor}

\mdfdefinestyle{fig}{skipabove=3pt,leftmargin=0pt,innerleftmargin=3pt,innertopmargin=3pt,skipbelow=3pt,rightmargin=0pt,innerrightmargin=3pt,innerbottommargin=3pt}

\definecolor{solarizedTextDark}{HTML}{002B36}
\definecolor{solarizedText}{HTML}{073642}
\definecolor{solarizedBaseDark}{HTML}{EEE8D5}
\definecolor{solarizedBase}{HTML}{FDF6E3}
\definecolor{sepiaBase}{HTML}{F6F0D6}
\definecolor{sepiaText}{HTML}{352F22}

\newcommand\cS{{\mathcal{S}}}
\newcommand\las{{\sf{s}}}
\newcommand\ie{{{\it i.e.,~}}}
\newcommand\eg{{{\it e.g.,~}}}

\usepackage{xr-hyper}
\usepackage[hidelinks]{hyperref} 

\title{Understanding Fundamental Tradeoffs in Nanomechanical Resonant Sensors}
 
\author{\vspace{0.5cm}{\large Alper~Demir}\\ \vspace{0.5cm}
{\small \textsf{Ko\c{c} University}}\\ \vspace{0.05cm}
{\small \textsf{Istanbul, Turkey}}\\ \vspace{0.1cm}
{\small \textsf{aldemir@ku.edu.tr}}\vspace{0.0cm}
}

\begin{document}

\maketitle
\flushbottom

\begin{abstract}
  Nanomechanical resonators are used as high performance detectors in
  a variety of applications such as mass spectrometry and atomic force
  microscopy. Initial emphasis in nanomechanical resonant sensor research 
  was on increasing the sensitivity to the level of a single molecule, 
  atom and beyond. On the other hand, there are applications where 
  the speed of detection is crucial, prompting recent works that emphasize 
  sensing schemes with improved time resolution. We 
  first develop a general modeling framework encompassing all resonator tracking 
  schemes currently in use, by extending recent previous work. 
  We then explore the fundamental
  trade-offs between accuracy and speed in three resonant sensor
  architectures, namely the feedback-free open-loop approach,
  positive-feedback based self-sustaining oscillator, and
  negative-feedback based frequency-locked loop scheme. We
  comparatively analyze them in a unified manner, clarify some
  misconceptions that seem to exist in the literature, and unravel their 
  speed versus accuracy characteristics. 
\end{abstract}

\section{Introduction}

Nanomechanical resonators are used as the core components in various
sensing, detection, spectroscopy and microscopy techniques, \eg in
mass spectrometry~\cite{chaste2012nanomechanical,naik2009towards,hanay2012single,Dominguez-Medina918,Ghadimi764} and atomic force microscopy~\cite{albrecht1991frequency,kobayashi2009frequency,adams2016harnessing,baykara2015noncontact,giessibl2019qplus}, Almost all of these
techniques are based on monitoring the resonance frequency shifts
arising from the interaction of the nanomechanical resonator with its
environment and the sample under study.

\begin{figure}
\centering
\includegraphics[width=0.6\columnwidth]{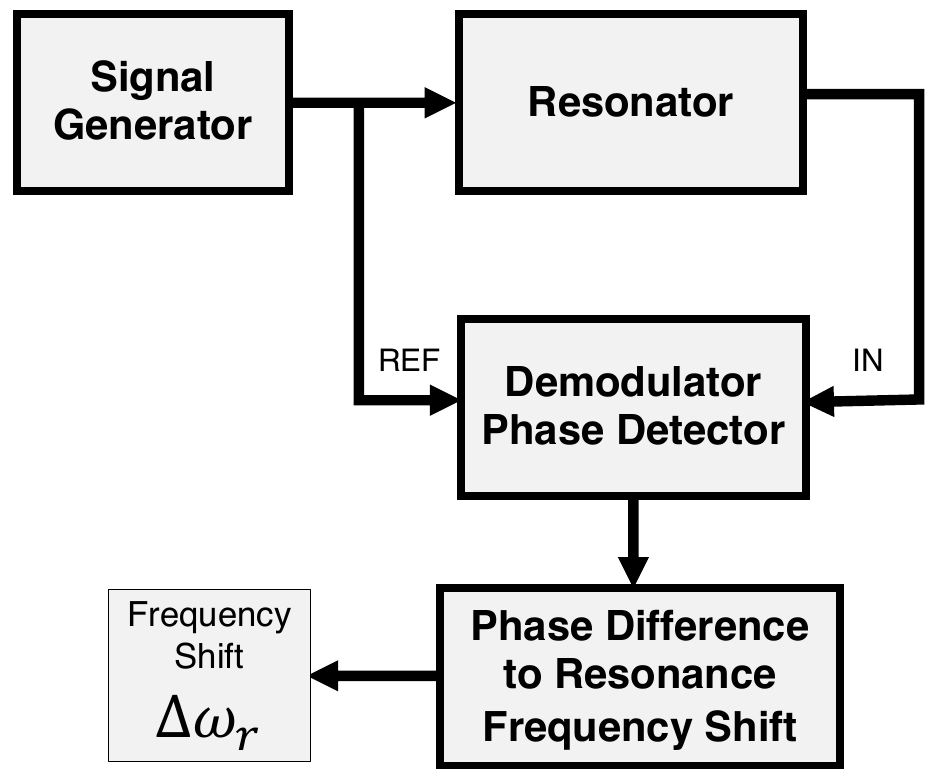}
\caption{Feedback-free resonator tracking (figure based on~\cite[Fig.~1]{demir2019fundamental})}
\label{fig:openloopblock}
\end{figure}

There are three prevalent schemes for measuring and tracking the
resonance frequency of a nanomechanical resonator. In the simplest
feedback-free ({\sc FF}) approach~\cite{albrecht1991frequency,sadeghi2020frequency}, shown in Fig.~\ref{fig:openloopblock}, the resonator is driven with a sinusoidal
signal at a constant amplitude, and a constant frequency close to the
resonance frequency. The vibratory motion of the resonator arising
from this excitation is transduced and demodulated in order to measure
its amplitude and phase difference with respect to the drive. Any
event of interest will result in a deviation in the measured response
amplitude and the phase difference, which can be used in conjunction with
the resonator characteristics in order to compute the induced
resonance frequency shift and other quantities such as mass changes
that relate to the cause. This feedback-free scheme has two major
disadvantages. The resonance frequency drifts over time, albeit at a
slow rate, due to phenomena other than the events of interest. If the
resonance frequency drifts too far away from the drive frequency, 
accurate measurement and computation of the frequency shift
becomes impossible due to degraded, flat resonator response. Even if
this drift problem is solved, the feedback-free scheme has an
inflexible trade-off characteristics between accuracy and speed that is
essentially dictated by the resonator characteristics, to be detailed
later.

\IEEEpubidadjcol

\begin{figure}
\centering
\includegraphics[width=0.65\columnwidth]{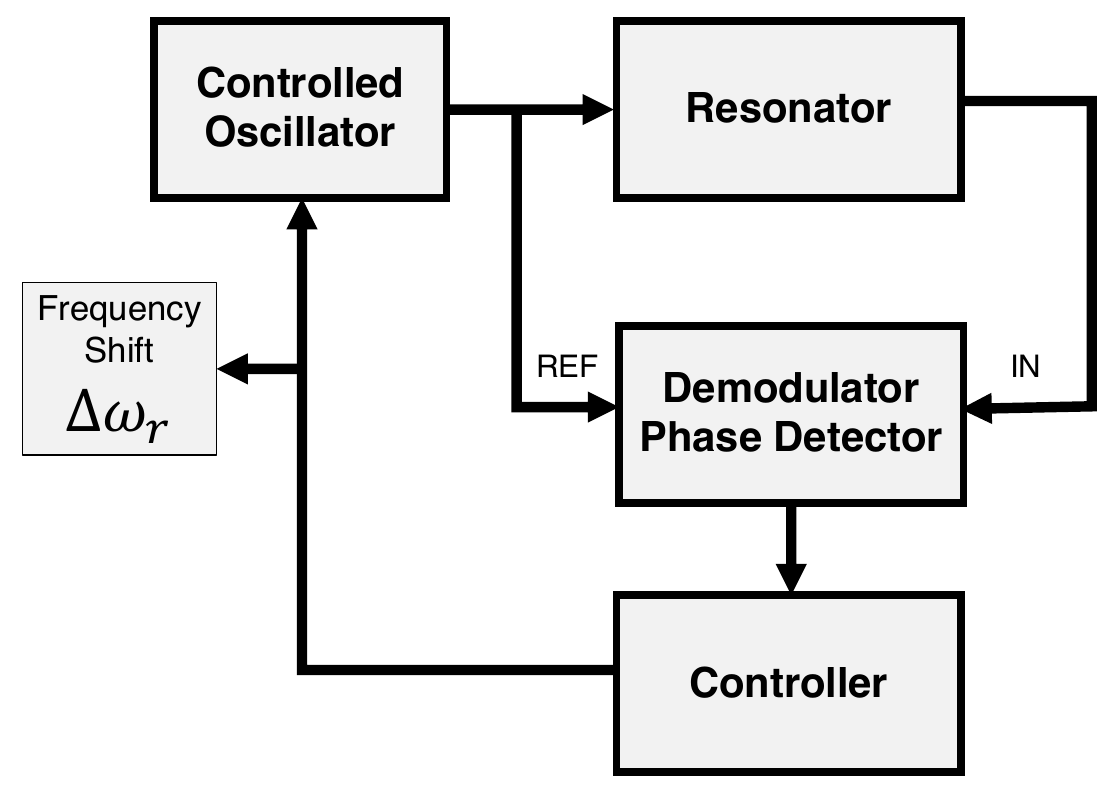}
\caption{Frequency-locked loop (figure based on~\cite[Fig.~1]{demir2019fundamental})}
\label{fig:fllblock}
\end{figure}

A {\em frequency-locked loop} ({\sc FLL})~\cite{naik2009towards,hanay2012single,olcum2015high,roy2018improving,demir2019fundamental} based resonator tracking
scheme, shown Fig.~\ref{fig:fllblock}, addresses the drift problem of the feedback-free approach,
by continually updating the frequency of the drive signal generator with a (negative) feedback control loop so that the resonator is
always driven at resonance. This solves the drift problem in the sense
that the resonator always operates in the region of its
characteristics that allows accurate frequency shift
discrimination. However, resonance frequency drifts due to other
nonideal phenomena may still interfere with the detection and tracking
of the events of interest that also result in a frequency
shift. Fortunately, in most applications, drift phenomena occur at a
much slower time scale, allowing the discrimination of the phenomena of
interest. The closed-loop {\sc FLL} scheme has another major advantage, 
offering flexibility in response speed. 
It is well established that resonators with higher quality factors offer
better sensitivity~\cite{demir2019fundamental}. 
However, a high quality factor also results in a
slow mechanical response dictating a lower detection speed.  On the
other hand, the speed with which an {\sc FLL} responds to frequency
shift events can be tailored based on the requirements of the
application. Importantly, this can be done in such a way so that the
{\sc FLL} response time is much shorter than, and in fact independent
of, the mechanical response time of the resonator. Unfortunately, this
added benefit does not come for free. As we show in detail later,
some of the sensitivity improvement obtained by using a resonator with
a higher quality factor has to be forfeited. On the other hand, the
{\sc FLL} dynamics can be designed to have a slower response when
compared with the mechanical response of the resonator, for an extra
sensitivity boost. This flexible speed-accuracy tradeoff
characteristics offered by an {\sc FLL} is most useful, but not very
well recognized and deeply understood in the literature. Finally, the {\sc
  FLL} scheme has a built-in utility for the measurement of the
resonance frequency via the feedback control signal that sets the
frequency of the drive signal generator.

\begin{figure}
\centering
\includegraphics[width=0.85\columnwidth]{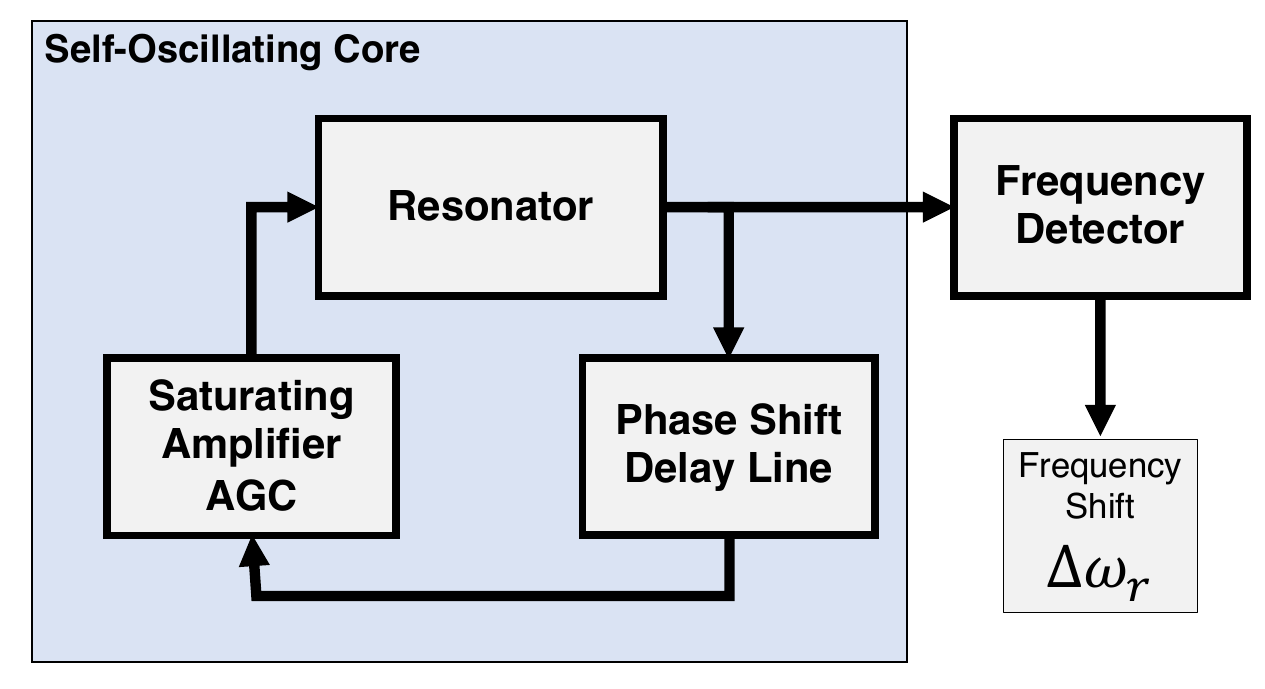}
\caption{Self-sustained oscillator (figure based on~\cite[Fig.~1]{demir2019fundamental})}
\label{fig:selfoscillatingblock}
\end{figure}

Another closed-loop method in use for nanomechanical resonant sensors
is the so-called {\em self-sustaining oscillator} ({\sc SSO}) scheme~\cite{albrecht1991frequency,durig1997dynamic,feng2008self}, 
shown in Fig.~\ref{fig:selfoscillatingblock}, where the resonator is
used as the frequency selective element in a classical
positive-feedback based oscillator architecture with amplification and
time delay (phase shift) in the loop. In the oscillator core, there is no external
signal generator. Instead, the periodic oscillation signal (with a
frequency close to the resonance frequency) is generated by the
marginal instability of the positive-feedback loop itself. Oscillators
of this sort require a mechanism for amplitude stabilization, which is
usually provided by a nonlinearity. For nanomechanical oscillators,
this nonlinearity can be implemented in the amplifier placed in the
loop, via gain saturation and automatic gain control ({\sc AGC}), as
in the {\sc FM-AFM}
scheme~\cite{albrecht1991frequency,durig1997dynamic} that is widely
used in atomic force microscopy. In this case, the resonator operates
in the linear regime. In other self-oscillating schemes, the resonator
is driven harder to operate in the nonlinear Duffing
regime~\cite{kenig2012optimal,villanueva2013surpassing,kenig2013phase,demir2018numerical}. 
Unlike in the case of {\sc FLL}, the self-oscillating scheme needs to be
augmented with a frequency measurement method such as a frequency
counter, an {\sc FM} detector based on Hilbert transforms or narrowband
{\sc IQ} demodulators, or a {\em phase-locked loop} ({\sc
  PLL})~\cite{albrecht1991frequency,durig1997dynamic}. A {\sc PLL} that
locks an independent, controllable signal generator to the oscillation
signal generated by the self-oscillating core has some advantages as
discussed in~\cite{durig1997dynamic}. We note the distinction between
the {\sc PLL} here and the {\sc FLL} discussed above, which are both
negative-feedback based control systems~\cite{demir2019fundamental}. 
The {\sc FLL} locks the
frequency of the signal generator directly to the resonator, whereas
the {\sc PLL} locks both the frequency and phase of the signal
generator to the oscillation signal produced by the positive-feedback
based self-oscillating core. In the {\sc FLL}, the resonator is driven
with an external signal generator, whereas in the self-oscillating
core, it is driven with an amplified and phase shifted version of the
signal itself generates via positive-feedback.  A hybrid scheme that
combines an {\sc FLL} with a self-oscillation loop was proposed
in~\cite{durig1997dynamic}, where the resonator is driven by a signal
that is obtained by combining its own amplified response with the
output of the signal generator in the {\sc PLL}.
The self-oscillating and the hybrid schemes described above offer similar 
advantages over the feedback-free approach. The resonator always operates 
at or near resonance in the linear regime, or at other desired operating 
points in the Duffing regime~\cite{kenig2012optimal,villanueva2013surpassing,kenig2013phase,demir2018numerical}. This solves the drift problem. The response 
time of a {\sc SSO} to an event of interest is also much shorter than, and in fact can be made independent
of, the mechanical response time of the resonator~\cite{albrecht1991frequency}, as in the case of an {\sc FLL}, 
which we derive in a rigorous manner in this paper. On the other hand, the {\sc SSO} 
scheme suffers from amplification of detection noise circulating 
in the positive feedback loop, resulting in an inferior accuracy (sensitivity) 
as compared with an {\sc FLL}, which we quantify precisely. 
Furthermore, the {\sc SSO} scheme also offers a flexible speed-accuracy 
trade-off characteristics, but not as versatile as the one in an {\sc FLL}.    

In Section~\ref{sec:theory}, we develop a unifying theory and models for resonators, basic configurations in resonant sensors and noise sources. The general modeling framework we develop in this paper encompasses all kinds of resonator tracking schemes, and extends our recent previous work in~\cite{demir2019fundamental} that considered only the {\sc FLL}. We describe the three standard resonator tracking schemes, {\sc FF}, {\sc FLL} and {\sc SSO} in Section~\ref{sec:schemes}, and perform a comparative characterization in Section~\ref{sec:analysis}. 

\section{Theory}
\label{sec:theory}

\subsection{Resonator model}
\label{sec:resmodel}

We consider a resonator that is modeled as a damped harmonic
oscillator as follows~\cite{hauer2013general} 
\begin{equation}
\dv[2]{}{t}\,x\pqty{t} + \frac{\omega_{\textsf{\tiny r}}}{Q}\,\dv{}{t}\,x\pqty{t} + \omega_{\textsf{\tiny r}}^2\,x\pqty{t} = \frac{F\pqty{t}}{m}
\label{eqn:resonator}
\end{equation}
where $x$ is the displacement, $m$ is the mass, $F\pqty{t}$ represents
a force excitation, $\omega_{\textsf{\tiny r}}$ is the resonance frequency, and $Q$ is
the quality factor. We express 
\begin{equation}
x\pqty{t} = {\cal R}\Bqty{s\pqty{t}\,e^{j\omega_{\textsf{\tiny o}} t}} =
\tfrac{1}{2}\bqty{s\pqty{t}\,e^{j\omega_{\textsf{\tiny o}} t}+s^*\pqty{t}\,e^{-j\omega_{\textsf{\tiny o}} t}}
\label{eqn:complexamp}
\end{equation}
where $\cdot^*$ denotes complex-conjugate, 
${\cal R}\Bqty{\cdot}$ is the real part operator. With a reasonably
high $Q$, $s\pqty{t}$ is the slowly varying complex amplitude of
$x\pqty{t}$. The carrier frequency $\omega_{\textsf{\tiny o}}$, while not necessarily
equal to the resonance frequency $\omega_{\textsf{\tiny r}}$, is chosen such that 
$\vqty{\omega_{\textsf{\tiny r}}-\omega_{\textsf{\tiny o}}} < \tfrac{\omega_{\textsf{\tiny r}}}{Q} \ll \omega_{\textsf{\tiny r}}$.
We substitute \eqref{eqn:complexamp} into \eqref{eqn:resonator} to
obtain
\begin{equation}
  \begin{aligned}
    \frac{1}{2} \bqty{ e^{j\omega_{\textsf{\tiny o}} t}\,{\tt \bf S}\pqty{t}+
    e^{-j\omega_{\textsf{\tiny o}} t}\,{\tt \bf S}^*\pqty{t} } =
    \frac{F\pqty{t}}{m}
  \end{aligned}
  \label{eqn:basebandexact}
\end{equation}
\begin{equation}
  {\tt \bf S}\pqty{t} =  \dv[2]{}{t}s +
      \pqty{\frac{\omega_{\textsf{\tiny r}}}{Q}+2j\omega_{\textsf{\tiny o}} }\dv{}{t}s
       + \pqty{{\omega_{\textsf{\tiny r}}^2-\omega_{\textsf{\tiny o}}^2}+ j\frac{\omega_{\textsf{\tiny r}}\,
          \omega_{\textsf{\tiny o}}}{Q} }s
\nonumber
\end{equation}
where the $t$ dependence of $s\pqty{t}$ was omitted for notational
simplicity.  We assume that $s\pqty{t}$ is slowly varying with a
maximum bandwidth of $\tfrac{w_r}{Q}$. Hence, the $t$ derivative of
$s\pqty{t}$ results in a maximum amplification of
$\tfrac{w_r}{Q}$. With
$\vqty{\omega_{\textsf{\tiny r}}-\omega_{\textsf{\tiny o}}} < \tfrac{\omega_{\textsf{\tiny r}}}{Q} \ll \omega_{\textsf{\tiny r}}$, the
terms above have the following amplification factor orders:
\begin{equation}
  \underbrace{\dv[2]{}{t}}_{{\tfrac{w_r^2}{Q^2}}} s +
      \underbrace{\frac{\omega_{\textsf{\tiny r}}}{Q}\dv{}{t}}_{\tfrac{w_r^2}{Q^2}}s+\underbrace{2j\omega_{\textsf{\tiny o}} \dv{}{t}}_{{\tfrac{w_r^2}{Q}}}s
       + \underbrace{\pqty{{\omega_{\textsf{\tiny r}}^2-\omega_{\textsf{\tiny o}}^2}+ j\frac{\omega_{\textsf{\tiny r}}\,
          \omega_{\textsf{\tiny o}}}{Q}}}_{{\tfrac{w_r^2}{Q}}}s
\nonumber
\end{equation}
The first two terms above are $Q$ times smaller than the last two
terms, and hence we neglect them and also use
$\omega_{\textsf{\tiny r}}^2-\omega_{\textsf{\tiny o}}^2\approx 2\omega_{\textsf{\tiny o}}\pqty{\omega_{\textsf{\tiny r}}-\omega_{\textsf{\tiny o}}}$
to obtain
\begin{equation}
  {\tt \bf S}\pqty{t} \approx  
      2j\omega_{\textsf{\tiny o}} \dv{}{t}s\pqty{t}
       + \pqty{2\omega_{\textsf{\tiny o}}\pqty{\omega_{\textsf{\tiny r}}-\omega_{\textsf{\tiny o}}}+ j\frac{\omega_{\textsf{\tiny r}}\,
          \omega_{\textsf{\tiny o}}}{Q} }s\pqty{t}
\label{eqn:Sapprox}
\end{equation}
$s\pqty{t}$ (and its derivative with respect to $t$) changes very
little over the cycle time of the resonator, \ie 
\begin{equation}
  {\tt \bf S}\pqty{t'}  \approx {\tt \bf S}\pqty{t}\quad \text{for} \quad t\leq
  t' \leq t+\tfrac{2\pi}{\omega_{\textsf{\tiny o}}}
  \label{eqn:slowamp}
\end{equation}
We then multiply both sides of \eqref{eqn:basebandexact} with
$\tfrac{\omega_{\textsf{\tiny o}}}{\pi}e^{-j\omega_{\textsf{\tiny o}} t}$ and integrate over a time interval of
$\tfrac{2\pi}{\omega_{\textsf{\tiny o}}}$ to obtain
\begin{equation}
  \begin{aligned}
    \frac{\omega_{\textsf{\tiny o}}}{2\pi} & \int_t^{t+\tfrac{2\pi}{\omega_{\textsf{\tiny o}}}}  \bqty{ {\tt \bf S}\pqty{t'}+
    e^{-j 2\omega_{\textsf{\tiny o}} t'}\,{\tt \bf S}^*\pqty{t'}} \dd t' \approx \\ &
  {\tt \bf S}\pqty{t} \approx \frac{1}{m}\frac{\omega_{\textsf{\tiny o}}}{\pi} \int_t^{t+\tfrac{2\pi}{\omega_{\textsf{\tiny o}}}}
    e^{-j\omega_{\textsf{\tiny o}} t'}  
    F\pqty{t'} \dd t'
  \end{aligned}
\end{equation}
We combine the above with \eqref{eqn:Sapprox}
\begin{equation}
  \begin{aligned}
      2j\omega_{\textsf{\tiny o}} \dv{}{t}s\pqty{t}
       + & \pqty{2\omega_{\textsf{\tiny o}}\pqty{\omega_{\textsf{\tiny r}}-\omega_{\textsf{\tiny o}}}+ j\frac{\omega_{\textsf{\tiny r}}\,
          \omega_{\textsf{\tiny o}}}{Q} }s\pqty{t}
    \approx \\ & \frac{1}{m}\frac{\omega_{\textsf{\tiny o}}}{\pi} \int_t^{t+\tfrac{2\pi}{\omega_{\textsf{\tiny o}}}}
    e^{-j\omega_{\textsf{\tiny o}} t'}  
    F\pqty{t'} \dd t'
  \end{aligned}
\end{equation}
multiply both sides with $-j\tfrac{Q}{\omega_{\textsf{\tiny r}}\,\omega_{\textsf{\tiny o}}}$
\begin{equation}
  \begin{aligned}
      \frac{2 Q}{\omega_{\textsf{\tiny r}}} \dv{}{t}s\pqty{t}
       + & \pqty{1 + j  \frac{2 Q}{\omega_{\textsf{\tiny r}}} \pqty{\omega_{\textsf{\tiny o}}-\omega_{\textsf{\tiny r}}}}s\pqty{t}
    \approx \\ & \frac{-j}{m}\frac{Q}{\omega_{\textsf{\tiny r}}}\frac{1}{\pi} \int_t^{t+\tfrac{2\pi}{\omega_{\textsf{\tiny o}}}}
    e^{-j\omega_{\textsf{\tiny o}} t'}  
    F\pqty{t'} \dd t'
  \end{aligned}
\end{equation}
and define $\tau_{\textsf{\tiny r}} = \tfrac{2 Q}{\omega_{\textsf{\tiny r}}}$ as the intrinsic
resonator time-constant to obtain
\begin{equation}
  \begin{aligned}
    \tau_{\textsf{\tiny r}} \dv{}{t}s\pqty{t}
    + & \bqty{1 + j \tau_{\textsf{\tiny r}} \pqty{\omega_{\textsf{\tiny o}}-\omega_{\textsf{\tiny r}}}}s\pqty{t}
    \approx \\ & \frac{-j}{m}\frac{Q}{\omega_{\textsf{\tiny r}}}\frac{1}{\pi} \int_t^{t+\tfrac{2\pi}{\omega_{\textsf{\tiny o}}}}
    e^{-j\omega_{\textsf{\tiny o}} t'}  
    F\pqty{t'} \dd t'
  \end{aligned}
  \label{eqn:resonatormodel}
\end{equation}

\subsection{Feedback-free driven resonator step response}
We choose the external force drive as 
\begin{equation}
F\pqty{t} = {\cal R}\Bqty{f\pqty{t}\,e^{j\omega_{\textsf{\tiny o}} t}} =
\tfrac{1}{2}\bqty{f\pqty{t}\,e^{j\omega_{\textsf{\tiny o}} t}+f^*\pqty{t}\,e^{-j\omega_{\textsf{\tiny o}} t}}
\label{eqn:drive}
\end{equation}
where $f\pqty{t}$ is a slowly varying complex amplitude, similar to
$s\pqty{t}$ in \eqref{eqn:complexamp}.  We substitute the above in
\eqref{eqn:resonatormodel}, assuming
\begin{equation}
f\pqty{t'} \approx f\pqty{t}\quad \text{for}
\quad t\leq t' \leq t+\tfrac{2\pi}{\omega_{\textsf{\tiny o}}}
\end{equation}
and obtain
\begin{equation}
  \begin{aligned}
      \tau_{\textsf{\tiny r}} \dv{}{t}s\pqty{t}
       +  \bqty{1 + j \tau_{\textsf{\tiny r}} \pqty{\omega_{\textsf{\tiny o}}-\omega_{\textsf{\tiny r}}}}s\pqty{t}
    = \frac{-j}{m}\frac{Q}{\omega_{\textsf{\tiny r}}\,\omega_{\textsf{\tiny o}}}f\pqty{t} 
  \end{aligned}
  \label{eqn:drivenresonatormodel}
\end{equation}
The above serves as a compact model for evaluating the resonator
response to various kinds of excitations.  We first choose
$f(t) = A_{\textsf{\tiny f}}e^{j\theta_{\textsf{\tiny f}}}$, where the amplitude $A_{\textsf{\tiny f}}$ and the phase
$\theta_{\textsf{\tiny f}}$ are time-invariant, which implies a sinusoidal drive
$F\pqty{t} = A_{\textsf{\tiny f}}\cos\pqty{\omega_{\textsf{\tiny o}} t + \theta_{\textsf{\tiny f}}}$. The steady-state
response of the resonator to this excitation can be computed by
setting the $t$ derivative in \eqref{eqn:drivenresonatormodel} to
zero as follows
\begin{equation}
  \begin{aligned}
    s\pqty{t} = 
    \frac{-j}{m}\frac{Q}{\omega_{\textsf{\tiny r}}\,\omega_{\textsf{\tiny o}}}\frac{A_{\textsf{\tiny f}}e^{j\theta_{\textsf{\tiny f}}}}{\bqty{1
        + j \tau_{\textsf{\tiny r}} \pqty{\omega_{\textsf{\tiny o}}-\omega_{\textsf{\tiny r}}}}}  = A_{\textsf{\tiny r}}e^{j\theta_{\textsf{\tiny r}}}
  \end{aligned}
  \label{eqn:drivenresonatorssresponse}
\end{equation}
where
\begin{equation}
  \begin{aligned}
    A_{\textsf{\tiny r}} &=& \frac{1}{m}\frac{Q}{\omega_{\textsf{\tiny r}}\,\omega_{\textsf{\tiny o}}}
    \frac{A_{\textsf{\tiny f}}}{\sqrt{1+\tau_{\textsf{\tiny r}}^2\pqty{\omega_{\textsf{\tiny o}}-\omega_{\textsf{\tiny r}}}^2}} \\
    \theta_{\textsf{\tiny r}} &=&
    {{\theta_{\textsf{\tiny f}}-\tfrac{\pi}{2}-\atan\pqty{\tau_{\textsf{\tiny r}}\pqty{\omega_{\textsf{\tiny o}}-\omega_{\textsf{\tiny r}}}}}}
  \end{aligned}
\label{eqn:ssampphase}
\end{equation}
implying a sinusoidal response
$x\pqty{t} = A_{\textsf{\tiny r}}\cos\pqty{\omega_{\textsf{\tiny o}} t + \theta_{\textsf{\tiny r}}}$. 

We next consider an excitation as follows
\begin{equation}
  f(t) = \left\{\begin{array}{ll} A_{\textsf{\tiny f}}\,e^{j\theta_{\textsf{\tiny f}}} & t < 0 \\ A_{\textsf{\tiny f}}\,e^{j\pqty{\Delta\omega_{\textsf{\tiny e}} t+\theta_{\textsf{\tiny f}}}} & t \geq 0 \end{array} \right.
  \label{eqn:drivenresonatortrainput}
\end{equation}
where the drive frequency is stepped at $t=0$ by $\Delta\omega_{\textsf{\tiny e}}$,
with $\Delta\omega_{\textsf{\tiny e}} < \tfrac{\omega_{\textsf{\tiny r}}}{Q}$. 
We evaluate the transient response of the
resonator due to the above excitation for $t\geq 0$, by setting the response
\begin{equation}
  \begin{aligned}
    s\pqty{t} = A_{\textsf{\tiny r}}\pqty{t}e^{j\pqty{\Delta\omega_{\textsf{\tiny e}} t+\theta_{\textsf{\tiny r}}\pqty{t}}}
  \end{aligned}
  \label{eqn:drivenresonatortraresponse}
\end{equation}
with the initial condition 
\begin{equation}
  \begin{aligned}
    s\pqty{0} = A_{\textsf{\tiny r}}\pqty{0}e^{j\theta_{\textsf{\tiny r}}\pqty{0}} =  \frac{-j}{m}\frac{Q}{\omega_{\textsf{\tiny r}}\,\omega_{\textsf{\tiny o}}}\frac{A_{\textsf{\tiny f}}e^{j\theta_{\textsf{\tiny f}}}}{\bqty{1
        + j \tau_{\textsf{\tiny r}} \pqty{\omega_{\textsf{\tiny o}}-\omega_{\textsf{\tiny r}}}}}
  \end{aligned}
  \label{eqn:drivenresonatortraresponseinitcond}
\end{equation}
with $A_{\textsf{\tiny r}}\pqty{0}$ and $\theta_{\textsf{\tiny r}}\pqty{0}$ as in
\eqref{eqn:ssampphase}.
We substitute \eqref{eqn:drivenresonatortrainput} and \eqref{eqn:drivenresonatortraresponse} in
\eqref{eqn:drivenresonatormodel} for $t\geq 0$, and after some
manipulations, we obtain 
  \begin{align}
  \nonumber
      \tau_{\textsf{\tiny r}} \dv{}{t} \bqty{A_{\textsf{\tiny r}}\pqty{t}e^{j\theta_{\textsf{\tiny r}}\pqty{t}}} 
       +  & \bqty{1 + j \tau_{\textsf{\tiny r}}
         \pqty{\omega_{\textsf{\tiny o}}+\Delta\omega_{\textsf{\tiny e}}-\omega_{\textsf{\tiny r}}}}{A_{\textsf{\tiny r}}\pqty{t}e^{j\theta_{\textsf{\tiny r}}\pqty{t}}}
       \\ & = \frac{-j}{m}\frac{Q}{\omega_{\textsf{\tiny r}}\,\omega_{\textsf{\tiny o}}} A_{\textsf{\tiny f}}e^{j\theta_{\textsf{\tiny f}}}
  \label{eqn:resonatormodelwithresfreqshift}
  \end{align}
We observe that this excitation scenario also
captures the case where the drive frequency is fixed, but the
resonance frequency suddenly changes,
\ie $\Delta w$ increase in the drive frequency may equivalently be
intrerpreted as a $\Delta w$
decrease in the resonance frequency. The solution of the above
differential equation with the initial condition in
\eqref{eqn:drivenresonatortraresponseinitcond} is given by
\eqref{eqn:drivenresonatortraresponsesoln}.
This response has a decay time constant $\tau_{\textsf{\tiny r}}$, but also
exhibits oscillatory behavior with beat frequency
$\omega_{\textsf{\tiny r}}-\omega_{\textsf{\tiny o}}-\Delta\omega_{\textsf{\tiny e}}$~\cite{albrecht1991frequency}.   

\begin{strip}
\noindent\makebox[\linewidth]{\rule{\textwidth}{1pt}} 
\begin{equation}
  \begin{aligned}
 A_{\textsf{\tiny r}}\pqty{t}e^{j\theta_{\textsf{\tiny r}}\pqty{t}} = 
 \frac{-j\: Q}{m\omega_{\textsf{\tiny r}}\,\omega_{\textsf{\tiny o}}} \bqty{
   \frac{A_{\textsf{\tiny f}}e^{j\theta_{\textsf{\tiny f}}}}
   {{1+ j \tau_{\textsf{\tiny r}} \pqty{\omega_{\textsf{\tiny o}}-\omega_{\textsf{\tiny r}}}}}  e^{-t/\tau_{\textsf{\tiny r}}}e^{j\pqty{\omega_{\textsf{\tiny r}}-\omega_{\textsf{\tiny o}}-\Delta\omega_{\textsf{\tiny e}}}t}
   + \frac{A_{\textsf{\tiny f}}e^{j\theta_{\textsf{\tiny f}}}}
   {{1 + j \tau_{\textsf{\tiny r}}\pqty{\omega_{\textsf{\tiny o}}+\Delta\omega_{\textsf{\tiny e}}-\omega_{\textsf{\tiny r}}}}} \pqty{1-e^{-t/\tau_{\textsf{\tiny r}}}e^{j\pqty{\omega_{\textsf{\tiny r}}-\omega_{\textsf{\tiny o}}-\Delta\omega_{\textsf{\tiny e}}}t}}} 
 \label{eqn:drivenresonatortraresponsesoln}
\end{aligned}
\end{equation}
\noindent\makebox[\linewidth]{\rule{\textwidth}{1pt}} 
\end{strip} 

\subsection{Feedback-free resonator phase response}
\label{sec:resphasemodel}

Due to its sharpness around the resonance frequency, the phase
response is preferred for measuring resonance
frequency shifts. To first order, the amplitude of the resonator
response around the resonance frequency can be approximated as
constant with $A_{\textsf{\tiny r}}\pqty{t} = A_{\textsf{\tiny rss}}$. We set $\omega_{\textsf{\tiny r}}=\omega_{\textsf{\tiny o}}$ and
\begin{equation}
A_{\textsf{\tiny r}}\pqty{t} = A_{\textsf{\tiny rss}} = \frac{Q\:A_{\textsf{\tiny f}}}{m\:\omega_{\textsf{\tiny r}}^2}\quad\text{for all}\quad t
\label{eqn:drivenssamp}
\end{equation}
in \eqref{eqn:resonatormodelwithresfreqshift} for
\begin{equation}
  \begin{aligned}
      \tau_{\textsf{\tiny r}} \dv{}{t} \bqty{e^{j\theta_{\textsf{\tiny r}}\pqty{t}}} 
       +  \pqty{1 + j \tau_{\textsf{\tiny r}}\Delta\omega_{\textsf{\tiny e}}}{e^{j\theta_{\textsf{\tiny r}}\pqty{t}}}
        = -j e^{j\theta_{\textsf{\tiny f}}}
  \end{aligned}
  \label{eqn:resonatormodelfrshiftforphase}
\end{equation}
with the initial condition $\theta_{\textsf{\tiny r}}\pqty{0} =
\theta_{\textsf{\tiny f}}-\tfrac{\pi}{2}$. We set $\theta_{\textsf{\tiny r}}\pqty{t} =
\theta_{\textsf{\tiny r}}\pqty{0} + \Delta\theta_{\textsf{\tiny r}}\pqty{t}$, with
$\Delta\theta_{\textsf{\tiny r}}\pqty{t}$ as the phase shift due to the resonance
frequency shift and subsititute into
\eqref{eqn:resonatormodelfrshiftforphase} for 
\begin{equation}
  \begin{aligned}
      \tau_{\textsf{\tiny r}} \dv{}{t} \bqty{e^{j\Delta\theta_{\textsf{\tiny r}}\pqty{t}}} 
       +  \pqty{1 + j \tau_{\textsf{\tiny r}}\Delta\omega_{\textsf{\tiny e}}}{e^{j\Delta\theta_{\textsf{\tiny r}}\pqty{t}}}
        = 1
  \end{aligned}
  \label{eqn:resonatormodelfrshiftforphasediff}
\end{equation}
Assuming a small phase shift, we use
$e^{j\Delta\theta_{\textsf{\tiny r}}\pqty{t}}\approx 1 + j\Delta\theta_{\textsf{\tiny r}}\pqty{t}$ in
the above to obtain
\begin{equation}
  \begin{aligned}
      j\tau_{\textsf{\tiny r}} \dv{}{t} {\Delta\theta_{\textsf{\tiny r}}\pqty{t}} 
       +  j \tau_{\textsf{\tiny r}}\Delta\omega_{\textsf{\tiny e}} + j\Delta\theta_{\textsf{\tiny r}}\pqty{t} -\tau_{\textsf{\tiny r}}\Delta\omega_{\textsf{\tiny e}}\Delta\theta_{\textsf{\tiny r}}\pqty{t} 
        = 0
  \end{aligned}
  \label{eqn:resonatormodelfrshiftforphasediffsimp}
\end{equation}
Since both $\Delta\omega_{\textsf{\tiny e}}$ and $\Delta\theta_{\textsf{\tiny r}}\pqty{t}$ are assumed to
be small, we neglect the second-order last term in the above for 
\begin{equation}
  \begin{aligned}
      \tau_{\textsf{\tiny r}} \dv{}{t} {\Delta\theta_{\textsf{\tiny r}}\pqty{t}} + \Delta\theta_{\textsf{\tiny r}}\pqty{t} = -\tau_{\textsf{\tiny r}}\Delta\omega_{\textsf{\tiny e}}
  \end{aligned}
  \label{eqn:resonatormodelfrshiftforphasediffsimpult}
\end{equation}
Thus, the phase response of the resonator to sudden resonance
frequency shifts of size $\Delta\omega_{\textsf{\tiny e}}$, \ie its {\it frequency step response} ({\sc
  FSTR}), can be obtained by simply applying a {\it phase step} of
size $\tau_{\textsf{\tiny r}}\Delta\omega_{\textsf{\tiny e}}$ at the input of a one-pole low-pass filter
based resonator model with transfer function
\begin{equation}
H_{\textsf{\tiny R}}\pqty{\las} =\frac{1}{1+\las\,\tau_{\textsf{\tiny r}}}  
\label{eqn:restranfun}
\end{equation}
A resonator model as above was also derived both in~\cite{demir2019fundamental} and~\cite{olcum2015high} and using two different techniques. We believe that the derivation we have presented above, a third alternative, is most general and systematic.  
The solution of \eqref{eqn:resonatormodelfrshiftforphasediffsimpult},
with initial condition $\Delta\theta_{\textsf{\tiny r}}\pqty{0}=0$, is 
\begin{equation}
      \Delta\theta_{\textsf{\tiny r}}\pqty{t} = -\tau_{\textsf{\tiny r}}\Delta\omega_{\textsf{\tiny e}}\pqty{1-e^{-t/\tau_{\textsf{\tiny r}}}} 
  \label{eqn:resonatormodelfrshiftforphasediffsimpultsoln}
\end{equation}

\subsection{Positive-feedback based oscillator model}

We consider a self-oscillating core, where an amplified and phase
shifted version of the resonator response is used to drive itself.
Based on the Barkhausen criterion, positive-feedback based oscillators
need to have an amplitude stabilization mechanism, that ensures a
consistent response at a desired level. This is provided by a
nonlinearity, \eg the Duffing nonlinearity of the resonator, a saturating
amplifier, or an {\sc AGC} loop. Furthermore,
in order to ensure proper positive-feedback and self-sustaining
oscillations, a phase condition needs to be fulfilled around the
loop, which may be achieved with a phase shifter, a delay line, or a
time-derivative in some cases.

We consider a self-oscillating core where the
resonator operates in the linear regime, and amplitude stabilization
is provided by a saturating amplifier. We set the excitation
$F\pqty{t}$ in \eqref{eqn:resonatormodel} as follows
\begin{equation}
  F\pqty{t} = e^{j\tfrac{\pi}{2}}\;h\pqty{x\pqty{t}}
  \label{eqn:selfoscfeedback}
\end{equation}
where $h\pqty{\cdot}$ above is a nonlinear odd-function, linear for small
values of its argument and with a saturating characteristics for large
input, \eg a $\tanh$ like function, or a polynomial with only first
and third order terms. The $\tfrac{\pi}{2}$ phase shift ensures proper
positive feedback and satisfaction of the Barkhausen criterion at the
resonance frequency, which may also be achieved with a time derivative
operating either on the argument or the return value of
$h\pqty{\cdot}$. 

We assume that the solution of the self-oscillating core is sinusoidal
or narrowband with the excitation as in \eqref{eqn:selfoscfeedback},
without higher-order harmonic components. For the feedback-free driven
case, this assumption is justified based on the linearity of the
resonator and the system and the fact that the drive itself is nearly
sinuosidal and narrowband. However, in the self-oscillating case, the
system is nonlinear due to the nonlinear function $h\pqty{\cdot}$ in
\eqref{eqn:selfoscfeedback}, even though the resonator itself is still
operating in the linear regime. That is, even if the argument
$x\pqty{t}$ of $h\pqty{\cdot}$ is narrowband, $h\pqty{x\pqty{t}}$ will
have higher-order harmonic components. However, these will be filtered
by the resonator, especially well, when $Q$ is large. 

We consider the integral in the right-hand-side of \eqref{eqn:resonatormodel}
with $F\pqty{t}$ as in \eqref{eqn:selfoscfeedback} and with
$x\pqty{t}$ in $\eqref{eqn:complexamp}$
\begin{equation}
  \begin{aligned}
 \int_t^{t+\tfrac{2\pi}{\omega_{\textsf{\tiny o}}}}
    e^{-j\omega_{\textsf{\tiny o}} t'}  
    h\pqty{\tfrac{1}{2}\bqty{s\pqty{t'}\,e^{j\omega_{\textsf{\tiny o}} t'}+s^*\pqty{t'}\,e^{-j\omega_{\textsf{\tiny o}} t'}}} \dd t'
  \end{aligned}
  \label{eqn:resonatormodelRHSselfosc}
\end{equation}
In the above operation, a nearly sinusoidal, narrowband signal with
carrier frequency $\omega_{\textsf{\tiny o}}$ is passed through a nonlinear function
$h\pqty{\cdot}$. The outcome will have narrowband components that are
centered around the harmonics of $\omega_{\textsf{\tiny o}}$. The integral above
extracts the narrowband signal component around the fundamental
$\omega_{\textsf{\tiny o}}$, assuming that $s\pqty{t}$ is slowly varying as before. 
When $h\pqty{\cdot}$ is an odd, memoryless nonlinear function as described
above, it can be shown that this component can be expressed as
\begin{equation}
  \begin{aligned}
    h_\textsf{\tiny D}\pqty{\vqty{s\pqty{t}}}\:s\pqty{t}
  \end{aligned}
  \label{eqn:resonatormodelRHSselfoscDF}
\end{equation}
where $h_\textsf{\tiny D}\pqty{\cdot}$ is a nonlinear function, related to but distinct
from $h\pqty{\cdot}$, known as a {\em
  describing function}~\cite{vander1968multiple}. $h_\textsf{\tiny D}\pqty{\vqty{s\pqty{t}}}$ 
essentially represents an amplitude-dependent gain. When
$h\pqty{\cdot}$ is a $\tanh$ like function, the corresponding
$h_\textsf{\tiny D}\pqty{\cdot}$ has a flat characteristics in the region where $h\pqty{\cdot}$ is
linear, and decays to small values when $h\pqty{\cdot}$
saturates. Hence, \eqref{eqn:resonatormodel} can be simplified as follows
\begin{equation}
  \begin{aligned}
    \tau_{\textsf{\tiny r}} \dv{}{t}s\pqty{t}
    +  & \bqty{1 + j \tau_{\textsf{\tiny r}} \pqty{\omega_{\textsf{\tiny o}}-\omega_{\textsf{\tiny r}}}}s\pqty{t}
   \\& = \frac{1}{m}\frac{Q}{\omega_{\textsf{\tiny r}}\,\omega_{\textsf{\tiny o}}} h_\textsf{\tiny D}\pqty{\vqty{s\pqty{t}}}\,s\pqty{t} 
  \end{aligned}
  \label{eqn:selfoscmodel}
\end{equation}
We assume a steady-state, sinusoidal solution to the above in the
form $s\pqty{t} = A_{\textsf{\tiny r}} e^{j\theta_{\textsf{\tiny r}}}$, with time-invariant $A_{\textsf{\tiny r}}$ and
$\theta_{\textsf{\tiny r}}$, and substitute in \eqref{eqn:selfoscmodel} to obtain 
\begin{equation}
  \begin{aligned} 
    \bqty{1 + j \tau_{\textsf{\tiny r}} \pqty{\omega_{\textsf{\tiny o}}-\omega_{\textsf{\tiny r}}}}A_{\textsf{\tiny r}} e^{j\theta_{\textsf{\tiny r}}} 
   = \frac{1}{m}\frac{Q}{\omega_{\textsf{\tiny r}}\,\omega_{\textsf{\tiny o}}} h_\textsf{\tiny D}\pqty{A_{\textsf{\tiny r}}}\,A_{\textsf{\tiny r}} e^{j\theta_{\textsf{\tiny r}}} 
  \end{aligned}
  \label{eqn:selfoscmodelsoln}
\end{equation}
We observe that the above equation does not impose any
restriction on the phase $\theta_{\textsf{\tiny r}}$. It is well-known that
autonomous oscillators have a {\em free} phase, determined
by the initial conditions, that may freely diffuse due to
noise~\cite{demirTCAS2000}. Without noise in the system, $\theta_{\textsf{\tiny r}}$
takes an arbitary but time-invariant value. We eliminate the terms in
\eqref{eqn:selfoscmodelsoln} that involve $\theta_{\textsf{\tiny r}}$
\begin{equation}
  \begin{aligned} 
    \bqty{1 + j \tau_{\textsf{\tiny r}} \pqty{\omega_{\textsf{\tiny o}}-\omega_{\textsf{\tiny r}}}}A_{\textsf{\tiny r}}
   = \frac{1}{m}\frac{Q}{\omega_{\textsf{\tiny r}}\,\omega_{\textsf{\tiny o}}} h_\textsf{\tiny D}\pqty{A_{\textsf{\tiny r}}}\,A_{\textsf{\tiny r}}
  \end{aligned}
  \label{eqn:selfoscmodelampsoln}
\end{equation}
$A_{\textsf{\tiny r}}$ is a real-valued solution of the above equation. The trivial
solution $A_{\textsf{\tiny r}}=0$ is in fact unstable due to positive feedback and net
gain in the loop. Any disturbance or nonzero initial condition will
kick the solution away from $A_{\textsf{\tiny r}}=0$. For
\eqref{eqn:selfoscmodelampsoln} to have a positive and real-valued
nontrivial solution $A_{\textsf{\tiny r}}$, both of the following conditions, obtained
from the real and imaginary parts of \eqref{eqn:selfoscmodelampsoln},
need to be satisfied
\begin{equation}
  \begin{aligned} 
    \omega_{\textsf{\tiny o}} &= \omega_{\textsf{\tiny r}} \\
   \frac{1}{m}\frac{Q}{\omega_{\textsf{\tiny r}}\,\omega_{\textsf{\tiny o}}} h_\textsf{\tiny D}\pqty{A_{\textsf{\tiny r}}} &= 1
  \end{aligned}
  \label{eqn:selfoscmodelampsolncond}
\end{equation}
Thus, the oscillation (carrier) frequency is in fact equal to the
resonance frequency $\omega_{\textsf{\tiny r}}$, ensured by the $\tfrac{\pi}{2}$ phase
shift introduced into the positive-feedback loop. $A_{\textsf{\tiny r}}$ is the unique
solution of the nonlinear equation in
\eqref{eqn:selfoscmodelampsolncond}.

\subsection{Self-oscillating core step response}
\label{sec:selfoscstepresp}

We evaluate the transient response of the self-oscillating core to a
sudden change in the resonance frequency, from $\omega_{\textsf{\tiny r1}}$ to
$\omega_{\textsf{\tiny r2}}$ at $t=0$.
We use the model in \eqref{eqn:selfoscmodel} for $t<0$, by choosing $\omega_{\textsf{\tiny o}} =
\omega_{\textsf{\tiny r}}=\omega_{\textsf{\tiny r1}}$. The steady-state solution for $t<0$ is then given by 
\begin{equation}
  \begin{aligned}
s\pqty{t} = A_{\textsf{\tiny r1}}\,e^{j\theta_{\textsf{\tiny r1}}}\quad\text{for}\quad t<0
\end{aligned}
\label{eqn:selfoscsolwr1}
\end{equation}
with arbitrary $\theta_{\textsf{\tiny r1}}$ determined by initial conditions at
$t=-\infty$, and $A_{\textsf{\tiny r1}}$ as the solution of
\begin{equation}
  \begin{aligned} 
   \frac{1}{m}\frac{Q}{\omega_{\textsf{\tiny r1}}^2} h_\textsf{\tiny D}\pqty{A_{\textsf{\tiny r1}}} &= 1
  \end{aligned}
  \label{eqn:selfoscsolwr1Ar1}
\end{equation}
and a carrier $e^{j\omega_{\textsf{\tiny r1}}t}$. 
We use the model in \eqref{eqn:selfoscmodel} for $t\geq 0$, by choosing $\omega_{\textsf{\tiny o}} =
\omega_{\textsf{\tiny r1}}$ and $\omega_{\textsf{\tiny r}}=\omega_{\textsf{\tiny r2}}$
\begin{equation}
  \begin{aligned}
    \tau_{\textsf{\tiny r}} \dv{}{t}s\pqty{t}
    +  & \bqty{1 + j \tau_{\textsf{\tiny r}} \pqty{\omega_{\textsf{\tiny r1}}-\omega_{\textsf{\tiny r2}}}}s\pqty{t}
   \\& = \frac{1}{m}\frac{Q}{\omega_{\textsf{\tiny r1}}\,\omega_{\textsf{\tiny r2}}} h_\textsf{\tiny D}\pqty{\vqty{s\pqty{t}}}\,s\pqty{t} 
  \end{aligned}
  \label{eqn:selfoscmodelwr2}
\end{equation}
with the initial condition in \eqref{eqn:selfoscsolwr1}. Assuming a
solution in the form $s\pqty{t} = A_{\textsf{\tiny r}}\pqty{t}\,e^{j\theta_{\textsf{\tiny r}}\pqty{t}}$ in the above, we obtain 
\begin{equation}
  \begin{aligned}
    \tau_{\textsf{\tiny r}} &\bqty{\dv{}{t}A_{\textsf{\tiny r}}\pqty{t}}e^{j\theta_{\textsf{\tiny r}}\pqty{t}}\\
    +  & \bqty{1 + j \tau_{\textsf{\tiny r}} \pqty{\dv{}{t}\theta_{\textsf{\tiny r}}\pqty{t} + \omega_{\textsf{\tiny r1}}-\omega_{\textsf{\tiny r2}}}}A_{\textsf{\tiny r}}\pqty{t}e^{j\theta_{\textsf{\tiny r}}\pqty{t}}
   \\ & = \frac{1}{m}\frac{Q}{\omega_{\textsf{\tiny r1}}\,\omega_{\textsf{\tiny r2}}} h_\textsf{\tiny D}\pqty{A_{\textsf{\tiny r}}\pqty{t}}\,A_{\textsf{\tiny r}}\pqty{t}e^{j\theta_{\textsf{\tiny r}}\pqty{t}} 
  \end{aligned}
  \label{eqn:selfoscmodelwr2Arthetar}
\end{equation}
All terms above have a common factor $e^{j\theta_{\textsf{\tiny r}}\pqty{t}}$ that can
be eliminated to obtain 
\begin{equation}
  \begin{aligned}
    \tau_{\textsf{\tiny r}}\dv{}{t}A_{\textsf{\tiny r}}\pqty{t}
    +  & \bqty{1 + j \tau_{\textsf{\tiny r}} \pqty{\dv{}{t}\theta_{\textsf{\tiny r}}\pqty{t} + \omega_{\textsf{\tiny r1}}-\omega_{\textsf{\tiny r2}}}}A_{\textsf{\tiny r}}\pqty{t}
   \\ & = \frac{1}{m}\frac{Q}{\omega_{\textsf{\tiny r1}}\,\omega_{\textsf{\tiny r2}}} h_\textsf{\tiny D}\pqty{A_{\textsf{\tiny r}}\pqty{t}}\,A_{\textsf{\tiny r}}\pqty{t} 
  \end{aligned}
  \label{eqn:selfoscmodelwr2Arthetarsimple}
\end{equation}
We note that $A_{\textsf{\tiny r}}\pqty{t}$ is a positive and real-valued quantity, the
real and imaginary parts of the above equation yields
\begin{equation}
  \begin{aligned}
    \dv{}{t}\theta_{\textsf{\tiny r}}\pqty{t} &=  \omega_{\textsf{\tiny r2}}-\omega_{\textsf{\tiny r1}}\\
    \tau_{\textsf{\tiny r}}\dv{}{t}A_{\textsf{\tiny r}}\pqty{t} +  A_{\textsf{\tiny r}}\pqty{t} &= \frac{1}{m}\frac{Q}{\omega_{\textsf{\tiny r1}}\,\omega_{\textsf{\tiny r2}}} h_\textsf{\tiny D}\pqty{A_{\textsf{\tiny r}}\pqty{t}}\,A_{\textsf{\tiny r}}\pqty{t} 
  \end{aligned}
  \label{eqn:selfoscmodelwr2Arthetarsimpleeqns}
\end{equation}
with the initial conditions $\theta_{\textsf{\tiny r}}\pqty{0}=\theta_{\textsf{\tiny r1}}$ and
$A_{\textsf{\tiny r}}\pqty{0} = A_{\textsf{\tiny r1}}$. The solution for the phase is given by 
\begin{equation}
  \theta_{\textsf{\tiny r}}\pqty{t} = \pqty{\omega_{\textsf{\tiny r2}}-\omega_{\textsf{\tiny r1}}}\:t + \theta_{\textsf{\tiny r1}} = \Delta\omega_{\textsf{\tiny e}}\:t + \theta_{\textsf{\tiny r1}} 
  \label{eqn:selfoscmodelphasestepresponse}
\end{equation}
The solution for the amplitude $A_{\textsf{\tiny r}}\pqty{t}$ can not be obtained
analytically, since the governing differential equation is
nonlinear. However, it can be approximated as follows.
$A_{\textsf{\tiny r2}}$, the new steady-state amplitude at the new resonance
frequency $\omega_{\textsf{\tiny r2}}$, is the solution of
\begin{equation}
  \begin{aligned} 
   \frac{1}{m}\frac{Q}{\omega_{\textsf{\tiny r1}}\,\omega_{\textsf{\tiny r2}}} h_\textsf{\tiny D}\pqty{A_{\textsf{\tiny r2}}} &= 1
  \end{aligned}
  \label{eqn:selfoscsolwr1Ar2}
\end{equation}
Let
\begin{equation}
A_{\textsf{\tiny r}}\pqty{t} =  A_{\textsf{\tiny r2}} + \Delta A_{\textsf{\tiny r}}\pqty{t} 
\label{eqn:ampexpansion}
\end{equation}
and assuming $A_{\textsf{\tiny r2}}-A_{\textsf{\tiny r1}}$ and $\Delta A_{\textsf{\tiny r}}\pqty{t}$ are small, use
the following approximation
\begin{equation}
  \begin{aligned}
    & \frac{1}{m}\frac{Q}{\omega_{\textsf{\tiny r1}}\,\omega_{\textsf{\tiny r2}}} h_\textsf{\tiny D}\pqty{A_{\textsf{\tiny r}}\pqty{t}}
    \approx \\ &\frac{1}{m}\frac{Q}{\omega_{\textsf{\tiny r1}}\,\omega_{\textsf{\tiny r2}}}
    h_\textsf{\tiny D}\pqty{A_{\textsf{\tiny r2}}} +
    \underbrace{\left.\frac{1}{m}\frac{Q}{\omega_{\textsf{\tiny r1}}\,\omega_{\textsf{\tiny r2}}}
      \dv{}{\bullet}h_\textsf{\tiny D}\pqty{\bullet}\right|_{\bullet=A_{\textsf{\tiny r2}}}}_{\gamma}\Delta A_{\textsf{\tiny r}}\pqty{t} \\ &
    \approx 1 + \gamma\,\Delta A_{\textsf{\tiny r}}\pqty{t}
  \end{aligned}
  \nonumber
  \label{eqn:selfoscmodelwr2Arapprox1}
\end{equation}
in \eqref{eqn:selfoscmodelwr2Arthetarsimpleeqns} for
\begin{equation}
  \begin{aligned}
    \tau_{\textsf{\tiny r}}\dv{}{t}\Delta A_{\textsf{\tiny r}}\pqty{t} +  A_{\textsf{\tiny r2}}+ \Delta A_{\textsf{\tiny r}}\pqty{t} &= \\
    \bqty{1 + \gamma\,\Delta A_{\textsf{\tiny r}}\pqty{t}} &\bqty{A_{\textsf{\tiny r2}} +\Delta A_{\textsf{\tiny r}}\pqty{t}} 
  \end{aligned}
  \label{eqn:selfoscmodelwr2Arsimpleeqn}
\end{equation}
Ignoring the second-order term in $\Delta A_{\textsf{\tiny r}}$, we obtain 
\begin{equation}
  \begin{aligned}
    \tau_{\textsf{\tiny r}}\dv{}{t}\Delta A_{\textsf{\tiny r}}\pqty{t} = 
    \gamma\,A_{\textsf{\tiny r2}}\,\Delta A_{\textsf{\tiny r}}\pqty{t} 
  \end{aligned}
  \label{eqn:selfoscmodelwr2Arsimpleeqn2}
\end{equation}
where $\gamma\,A_{\textsf{\tiny r2}}$ is a dimensionless, negative valued quantity.
Based on the above equation, and given that $A_{\textsf{\tiny r}}\pqty{0} = A_{\textsf{\tiny r1}}$, we
obtain
\begin{equation}
  \begin{aligned}
  A_{\textsf{\tiny r}}\pqty{t} = A_{\textsf{\tiny r2}} + \pqty{A_{\textsf{\tiny r1}}-A_{\textsf{\tiny r2}}} \:e^{\gamma\,A_{\textsf{\tiny r2}}\,\tfrac{t}{\tau_{\textsf{\tiny r}}}}
  \end{aligned}
  \label{eqn:selfoscmodelampstepresponse}
\end{equation}

\subsection{Thermomechanical noise in the resonator model}

The {\em thermomechanical noise} of the resonator can be modeled as a
white noise source at the drive input $F\pqty{t}$ in
\eqref{eqn:resonator}, with a (two-sided) spectral density
\begin{equation}
\cS_{\textsf{\tiny thm}}\pqty{\omega} = 2\,m\,\frac{\omega_{\textsf{\tiny r}}}{Q}\,k_{\textsf{\tiny B}}\,T
\label{eqn:thmnoisePSD} 
\end{equation}
where $k_{\textsf{\tiny B}}$ is Boltzmann's constant, and $T$ is
temperature~\cite{ekinci2004ultimate}. In order to incorporate this
noise source into the model that was developed in
Section~\ref{sec:resmodel}, we use a complex amplitude
representation~\cite{benedetto1999principles} as follows
\begin{equation}
  \begin{aligned}
  N_{\textsf{\tiny thm}}\pqty{t} & = {\cal R}\Bqty{n_{\textsf{\tiny thm}}\pqty{t}\,e^{j\omega_{\textsf{\tiny o}} t}} 
\\ & = \tfrac{1}{2}\bqty{n_{\textsf{\tiny thm}}\pqty{t}\,e^{j\omega_{\textsf{\tiny o}} t}+n^*_{\textsf{\tiny thm}}\pqty{t}\,e^{-j\omega_{\textsf{\tiny o}} t}}
\end{aligned}
\label{eqn:thmnoisebb}
\end{equation}
with
\begin{equation}
  \begin{aligned}
  n_{\textsf{\tiny thm}}\pqty{t} = n_{\textsf{\tiny thm}\cal{\tiny R}}\pqty{t} + j\:n_{\textsf{\tiny thm}\cal{\tiny I}}\pqty{t} 
\end{aligned}
\label{eqn:thmnoisebbreim}
\end{equation}
where the real and imaginary parts $n_{\textsf{\tiny thm}\cal{\tiny R}}\pqty{t}$ and
$n_{\textsf{\tiny thm}\cal{\tiny I}}\pqty{t}$ are both white noise sources, uncorrelated with
each other, and with twice the spectral
density~\cite{benedetto1999principles} in \eqref{eqn:thmnoisePSD}:
\begin{equation}
\cS_{\textsf{\tiny thm}{\cal R}}\pqty{\omega} = \cS_{\textsf{\tiny thm}{\cal I}}\pqty{\omega} = 4\,m\,\frac{\omega_{\textsf{\tiny r}}}{Q}\,k_{\textsf{\tiny B}}\,T
\label{eqn:thmnoisePSDreim} 
\end{equation}
The resonator model with thermomechanical noise, obtained from
\eqref{eqn:drivenresonatormodel}, is then given by
\begin{equation}
  \begin{aligned}
      \tau_{\textsf{\tiny r}} \dv{}{t}s\pqty{t}
       +  & \bqty{1 + j \tau_{\textsf{\tiny r}} \pqty{\omega_{\textsf{\tiny o}}-\omega_{\textsf{\tiny r}}}}s\pqty{t}
    = \\ & \frac{-j}{m}\frac{Q}{\omega_{\textsf{\tiny r}}\,\omega_{\textsf{\tiny o}}}\bqty{f\pqty{t}+n_{\textsf{\tiny thm}}\pqty{t}} 
  \end{aligned}
  \label{eqn:drivenresonatormodelwiththmnoise}
\end{equation}
The noise source can also be incorporated into the phase domain model
developed in Section~\ref{sec:resphasemodel} by modifying \eqref{eqn:resonatormodelfrshiftforphasediff}
(with the frequency shift $\Delta\omega_{\textsf{\tiny e}}$ set to zero) as follows 
\begin{equation}
  \begin{aligned}
      \tau_{\textsf{\tiny r}} \dv{}{t} \bqty{e^{j\Delta\theta_{\textsf{\tiny r}}\pqty{t}}} 
       +  {e^{j\Delta\theta_{\textsf{\tiny r}}\pqty{t}}}
        = 1 + \tfrac{1}{A_{\textsf{\tiny f}}}n_{\textsf{\tiny thm}}\pqty{t}
  \end{aligned}
  \label{eqn:resonatormodelfrshiftforphasediffwiththmnoise}
\end{equation}
We note here that any phase shift (possibly time-varying) applied to $n_{\textsf{\tiny thm}}\pqty{t}$ does not
change the stochastic properties of its real and imaginary parts,
\ie, $n_{\textsf{\tiny thm}}\pqty{t}$ and $e^{j\theta\pqty{t}}n_{\textsf{\tiny thm}}\pqty{t}$ are
stochastically equivalent. Thus, we ignore any such phase shifts
that are applied to $n_{\textsf{\tiny thm}}\pqty{t}$. 
Subsequently, the noise term in the above dictates a noise term in
\eqref{eqn:resonatormodelfrshiftforphasediffsimpult} as follows
\begin{equation}
  \begin{aligned}
      \tau_{\textsf{\tiny r}} \dv{}{t} {\Delta\theta_{\textsf{\tiny r}}\pqty{t}} +
      \Delta\theta_{\textsf{\tiny r}}\pqty{t} & = \tfrac{1}{A_{\textsf{\tiny f}}}n_{\textsf{\tiny thm}{\cal
            I}}\pqty{t}  \\ & = 
      \underbrace{\tfrac{1}{A_{\textsf{\tiny rss}}}\tfrac{Q}{m\,\omega^2_{r}}
    \:n_{\textsf{\tiny thm}\cal{I}}\pqty{t}}_{\theta_{\textsf{\tiny th}}\pqty{t}} 
  \end{aligned}
  \label{eqn:resonatormodelfrshiftforphasediffsimpultwiththmnoise}
\end{equation}
where we used \eqref{eqn:drivenssamp}.  Thus, the phase deviation
$\Delta\theta_{\textsf{\tiny r}}\pqty{t}$ due to thermomechanical noise of the
resonator can be modeled with a white noise process
$\theta_{\textsf{\tiny th}}\pqty{t}$, at the input of the one-pole filter represented
by the transfer function in \eqref{eqn:restranfun}.
$\theta_{\textsf{\tiny th}}\pqty{t}$ is given by 
\begin{equation}
  \theta_{\textsf{\tiny th}}\pqty{t} = \frac{1}{A_{\textsf{\tiny rss}}}\frac{Q}{m\,\omega^2_{r}} \:n_{\textsf{\tiny thm}\cal{I}}\pqty{t}
\label{eqn:thetan} 
\end{equation}

\subsection{Detection  noise}

The thermomechanical noise of the resonator is the most fundamental,
unavoidable noise source. However, there are other noise sources that
need to be taken into account, for instance, the ones that are
introduced during the transduction of the mechanical displacement of
the resonator into an optical and/or electrical signal and in the subsequent
amplification in the electrical domain. Regardless of the actual
physical source, we lump all of these noise sources into one, represent them as an additive, {\em detection} noise source at the
output of the resonator (added to $s\pqty{t}$, at the input of any
electrical amplifier), and define it as follows
\begin{equation}
  \begin{aligned}
n_{\textsf{\tiny d}}\pqty{t} = {\cal K}_{\textsf{\tiny d}}\,\frac{Q}{m\,\omega_{\textsf{\tiny r}}^2}\,n_{\textsf{\tiny thmD}}\pqty{t} = n_{\textsf{\tiny d}\cal{R}}\pqty{t} + j\:n_{\textsf{\tiny d}\cal{I}}\pqty{t}
  \end{aligned}
  \label{eqn:detectionnoiseadd}
\end{equation}
representing a baseband equivalent, complex-valued white noise source,
as in \eqref{eqn:thmnoisebbreim}.  In \eqref{eqn:detectionnoiseadd},
the dimensionless factor ${\cal K}_{\textsf{\tiny d}}$ determines whether the
thermomechanical noise is resolved above the detection noise
background (${\cal K}_{\textsf{\tiny d}}<1$) or buried in it (${\cal K}_{\textsf{\tiny d}}>1$), as
implied by \eqref{eqn:drivenresonatormodelwiththmnoise}. The value of
${\cal K}_{\textsf{\tiny d}}$ can be easily determined experimentally when
${\cal K}_{\textsf{\tiny d}}<1$, by simply measuring the spectrum of the transduced
resonator displacement signal in the electrical domain in an open-loop
configuration with no excitation on the resonator~\cite{sadeghi2020frequency}. The (square root)
of the ratio of the spectrum level of the (white) background to the
level of the peak at the resonance frequency yields ${\cal K}_{\textsf{\tiny d}}$~\cite[Fig.~1(b)]{sadeghi2020frequency}.
$n_{\textsf{\tiny thmD}}\pqty{t}$ in \eqref{eqn:detectionnoiseadd} is a noise source
that is uncorrelated with $n_{\textsf{\tiny thm}}\pqty{t}$ but has the same
stochastic characteristics, \ie spectral density.   

The phase deviation $\theta_{\textsf{\tiny d}}\pqty{t}$ of a driven resonator due to detection noise, in an open-loop
configuration, can be modeled by adding detection
noise to the resonator response as follows
\begin{equation}
  \begin{aligned}
    A_{\textsf{\tiny rss}} +  n_{\textsf{\tiny d}\cal{\tiny R}}\pqty{t} + j\,n_{\textsf{\tiny d}\cal{I}}\pqty{t} = & \bqty{A_{\textsf{\tiny rss}} +\Delta A_{\textsf{\tiny r}}\pqty{t}} \,e^{j\theta_{\textsf{\tiny d}}\pqty{t}}\\
\approx & A_{\textsf{\tiny rss}} + \Delta A_{\textsf{\tiny r}}\pqty{t} +
j\,A_{\textsf{\tiny rss}}\,\theta_{\textsf{\tiny d}}\pqty{t}
\end{aligned}
\nonumber
\end{equation}
where we used $e^{j\theta_{\textsf{\tiny d}}\pqty{t}}\approx 1 +
\theta_{\textsf{\tiny d}}\pqty{t}$ and ignored the second-order noise term
$\Delta A_{\textsf{\tiny r}}\pqty{t}\,\theta_{\textsf{\tiny d}}\pqty{t}$. We identify 
\begin{equation}
  \begin{aligned}
    \theta_{\textsf{\tiny d}}\pqty{t} = \frac{n_{\textsf{\tiny d}\cal{I}}\pqty{t}}{A_{\textsf{\tiny rss}}} =
    \frac{{\cal K}_{\textsf{\tiny d}}}{A_{\textsf{\tiny rss}}}\,\frac{Q}{m\,\omega_{\textsf{\tiny r}}^2}\,n_{\textsf{\tiny thmD}{\cal I}}\pqty{t}
  \end{aligned}
  \label{eqn:detectionnoiseinoutput}
\end{equation}
The white noise process $\theta_{\textsf{\tiny d}}\pqty{t}$ above, representing detection
noise, is to be added to the
{\em output} of the one-pole filter in \eqref{eqn:restranfun}, whereas
$\theta_{\textsf{\tiny th}}\pqty{t}$ in \eqref{eqn:thetan}, representing
thermomechanical noise, is introduced at its {\em input}.  

\subsection{Noise in the self-oscillating core}

The thermomechanical noise of the resonator and detection noise are introduced into the
model of the self-oscillating core in \eqref{eqn:selfoscmodel}
\begin{equation}
  \begin{aligned}
    \tau_{\textsf{\tiny r}} &\dv{}{t}s\pqty{t}
    +   \bqty{1 + j \tau_{\textsf{\tiny r}} \pqty{\omega_{\textsf{\tiny o}}-\omega_{\textsf{\tiny r}}}}s\pqty{t}
   \\& = \frac{1}{m}\frac{Q}{\omega_{\textsf{\tiny r}}\,\omega_{\textsf{\tiny o}}}
   \bigg[h_\textsf{\tiny D}\pqty{\vqty{s\pqty{t}}}\,\bqty{s\pqty{t}+n_{\textsf{\tiny d}}\pqty{t}}+n_{\textsf{\tiny thm}}\pqty{t}\bigg] 
  \end{aligned}
  \label{eqn:selfoscmodelwithnoise}
\end{equation}
as in \eqref{eqn:drivenresonatormodelwiththmnoise}.
Above, we did not include $n_{\textsf{\tiny d}}\pqty{t}$ in the argument of
$h_\textsf{\tiny D}\pqty{\cdot}$, ignoring the second-order, detection noise effect in
determining the amplitude-dependent gain value for the saturating amplifier.
$n_{\textsf{\tiny d}}\pqty{t}$ in \eqref{eqn:selfoscmodelwithnoise} represents both
the noise introduced during the transduction as well as the noise of the
amplifier (in an input-referred manner). $n_{\textsf{\tiny d}}\pqty{t}$ is amplified
along with the signal component $s\pqty{t}$ and fed back to drive the
resonator. As before, the phase shift or time delay introduced in the
self-oscillating core does not change the stochastic properties of
$n_{\textsf{\tiny d}}\pqty{t}$. 

In order to evaluate the impact of noise on the
amplitude and phase of the signal generated by the self-oscillating
core, we modify \eqref{eqn:selfoscmodelwr2Arthetarsimple} as follows,
with $\omega_{\textsf{\tiny r}}=\omega_{\textsf{\tiny r1}}=\omega_{\textsf{\tiny r2}}$,
\begin{equation}
  \begin{aligned}
    \tau_{\textsf{\tiny r}}&\dv{}{t}A_{\textsf{\tiny r}}\pqty{t}
    +   \bqty{1 + j \tau_{\textsf{\tiny r}} {\dv{}{t}\theta_{\textsf{\tiny r}}\pqty{t} }}A_{\textsf{\tiny r}}\pqty{t}
   \\ & = \frac{1}{m}\frac{Q}{\omega^2_{r}}
   \bigg[h_\textsf{\tiny D}\pqty{A_{\textsf{\tiny r}}\pqty{t}}\,\bqty{A_{\textsf{\tiny r}}\pqty{t}+n_{\textsf{\tiny d}}\pqty{t}}+n_{\textsf{\tiny thm}}\pqty{t}\bigg]
  \end{aligned}
  \label{eqn:selfoscmodelwr2Arthetarsimplewiththmnoise}
\end{equation}
by ignoring any phase shifts that are applied to $n_{\textsf{\tiny thm}}\pqty{t}$ and
$n_{\textsf{\tiny d}}\pqty{t}$ in deriving
\eqref{eqn:selfoscmodelwr2Arthetarsimplewiththmnoise} from
\eqref{eqn:selfoscmodelwithnoise}. Seperating the real/imaginary
parts of the above equation results in
\begin{equation}
  \begin{aligned}
    \tau_{\textsf{\tiny r}} &A_{\textsf{\tiny r}}\pqty{t}\:\dv{}{t}\theta_{\textsf{\tiny r}}\pqty{t} = \\ & \frac{1}{m}\frac{Q}{\omega^2_{r}}
    \bqty {h_\textsf{\tiny D}\pqty{A_{\textsf{\tiny r}}\pqty{t}}\,n_{\textsf{\tiny d}\cal{I}}\pqty{t} + n_{\textsf{\tiny thm}\cal{I}}\pqty{t} }\\
    \tau_{\textsf{\tiny r}}&\dv{}{t}  A_{\textsf{\tiny r}}\pqty{t}  +  A_{\textsf{\tiny r}}\pqty{t} =\\
    & \frac{1}{m}\frac{Q}{\omega^2_{r}}
    \bigg[h_\textsf{\tiny D}\pqty{A_{\textsf{\tiny r}}\pqty{t}}\,\bqty{A_{\textsf{\tiny r}}\pqty{t}+n_{\textsf{\tiny d}\cal{R}}\pqty{t}} + n_{\textsf{\tiny thm}\cal{R}}\pqty{t}\bigg]
  \end{aligned}
  \label{eqn:selfoscmodelwr2Arthetarsimpleeqnswiththmnoise}
\end{equation}
The above equations can be simplified, nothing that there is a one way
coupling between them (from $A_{\textsf{\tiny r}}\pqty{t}$ to $\theta_{\textsf{\tiny r}}\pqty{t}$, but
not in the reverse direction) and by expressing $\theta_{\textsf{\tiny r}}\pqty{t} =
\theta_{\textsf{\tiny rss}} + \Delta\theta_{\textsf{\tiny r}}\pqty{t}$ and  $A_{\textsf{\tiny r}}\pqty{t} = A_{\textsf{\tiny rss}}
+ \Delta A_{\textsf{\tiny r}}\pqty{t}$ with the steady-state, noiseless $A_{\textsf{\tiny rss}}$ as the
solution of
\begin{equation}
  \begin{aligned} 
   \frac{1}{m}\frac{Q}{\omega^2_{r}} h_\textsf{\tiny D}\pqty{A_{\textsf{\tiny rss}}} = 1
  \end{aligned}
  \label{eqn:selfoscsolwrArss}
\end{equation}
$\theta_{\textsf{\tiny rss}}$ is an arbitrary but constant phase determined by
the initial conditions.  We proceed as in
\eqref{eqn:ampexpansion}-\eqref{eqn:selfoscmodelwr2Arsimpleeqn2} to
obtain 
\begin{equation}
  \begin{aligned}
    \tau_{\textsf{\tiny r}} &\bqty{A_{\textsf{\tiny rss}} +
      \Delta A_{\textsf{\tiny r}}\pqty{t}} \:\dv{}{t}\Delta\theta_{\textsf{\tiny r}}\pqty{t} = \\
    &\frac{1}{m}\frac{Q}{\omega^2_{r}}
    \bqty{h_\textsf{\tiny D}\pqty{A_{\textsf{\tiny rss}}}\,n_{\textsf{\tiny d}\cal{I}}\pqty{t} + \:n_{\textsf{\tiny thm}\cal{I}}\pqty{t}} \\
    \tau_{\textsf{\tiny r}}&\dv{}{t}\Delta A_{\textsf{\tiny r}}\pqty{t} -
    \gamma\,A_{\textsf{\tiny rss}}\,\Delta A_{\textsf{\tiny r}}\pqty{t}= \\
    &\frac{1}{m}\frac{Q}{\omega^2_{r}}
    \bqty{h_\textsf{\tiny D}\pqty{A_{\textsf{\tiny rss}}}\,n_{\textsf{\tiny d}\cal{R}}\pqty{t} + \:n_{\textsf{\tiny thm}\cal{R}}\pqty{t}}
  \end{aligned}
  \label{eqn:selfoscmodelwr2Arthetarsimplereqnswiththmnoise}
\end{equation}
The amplitude deviation $\Delta A_{\textsf{\tiny r}}\pqty{t}$ and the
phase deviation $\Delta\theta_{\textsf{\tiny r}}\pqty{t}$ are due to 
noise. Due to amplitude stabilization in the self-oscillating core,
$\Delta A_{\textsf{\tiny r}}\pqty{t}$ stays small compared with $A_{\textsf{\tiny rss}}$. Based on the
second equation above, $\Delta A_{\textsf{\tiny r}}\pqty{t}$ is essentially the output
of a one-pole (with time constant $\tfrac{-\tau_{\textsf{\tiny r}}}{\gamma\,A_{\textsf{\tiny rss}}}$)
low-pass filter, with the input as a white noise process.  
We thus approximate the first equation above using $A_{\textsf{\tiny rss}} + \Delta
A_{\textsf{\tiny r}}\pqty{t} \approx A_{\textsf{\tiny rss}}$, and obtain
\begin{align}
  \nonumber
    \dv{}{t}\Delta\theta_{\textsf{\tiny r}}\pqty{t} = & \frac{1}{\tau_{\textsf{\tiny r}}}
    \frac{1}{ A_{\textsf{\tiny rss}}}\frac{Q}{m\,\omega^2_{r}}
    \bqty{h_\textsf{\tiny D}\pqty{A_{\textsf{\tiny rss}}}\,n_{\textsf{\tiny d}\cal{I}}\pqty{t} +
                                       \:n_{\textsf{\tiny thm}\cal{I}}\pqty{t}} \\ =
& \frac{1}{\tau_{\textsf{\tiny r}}} \bqty{ \frac{Q}{m\,\omega^2_{r}}h_\textsf{\tiny D}\pqty{A_{\textsf{\tiny rss}}}\,\theta_{\textsf{\tiny d}}\pqty{t}+ \theta_{\textsf{\tiny th}}\pqty{t}}
\label{eqn:selfoscmodelwr2Arthetarsimplesteqnswiththmnoise}
\end{align}
The above equation implies that the phase deviation
$\Delta\theta_{\textsf{\tiny r}}\pqty{t}$ is the (time) integral of a white noise
process, thus in the form of a Brownian motion (random walk). The fact
that the phase noise of autonomous oscillators has a Browinan motion
nature is well established~\cite{demirTCAS2000}. 

\section{Resonator tracking schemes}
\label{sec:schemes}

\subsection{Feedback-free (FF) resonator tracking}
\label{sec:ffopenloop}

\begin{figure}
\centering
\includegraphics[width=1.0\columnwidth]{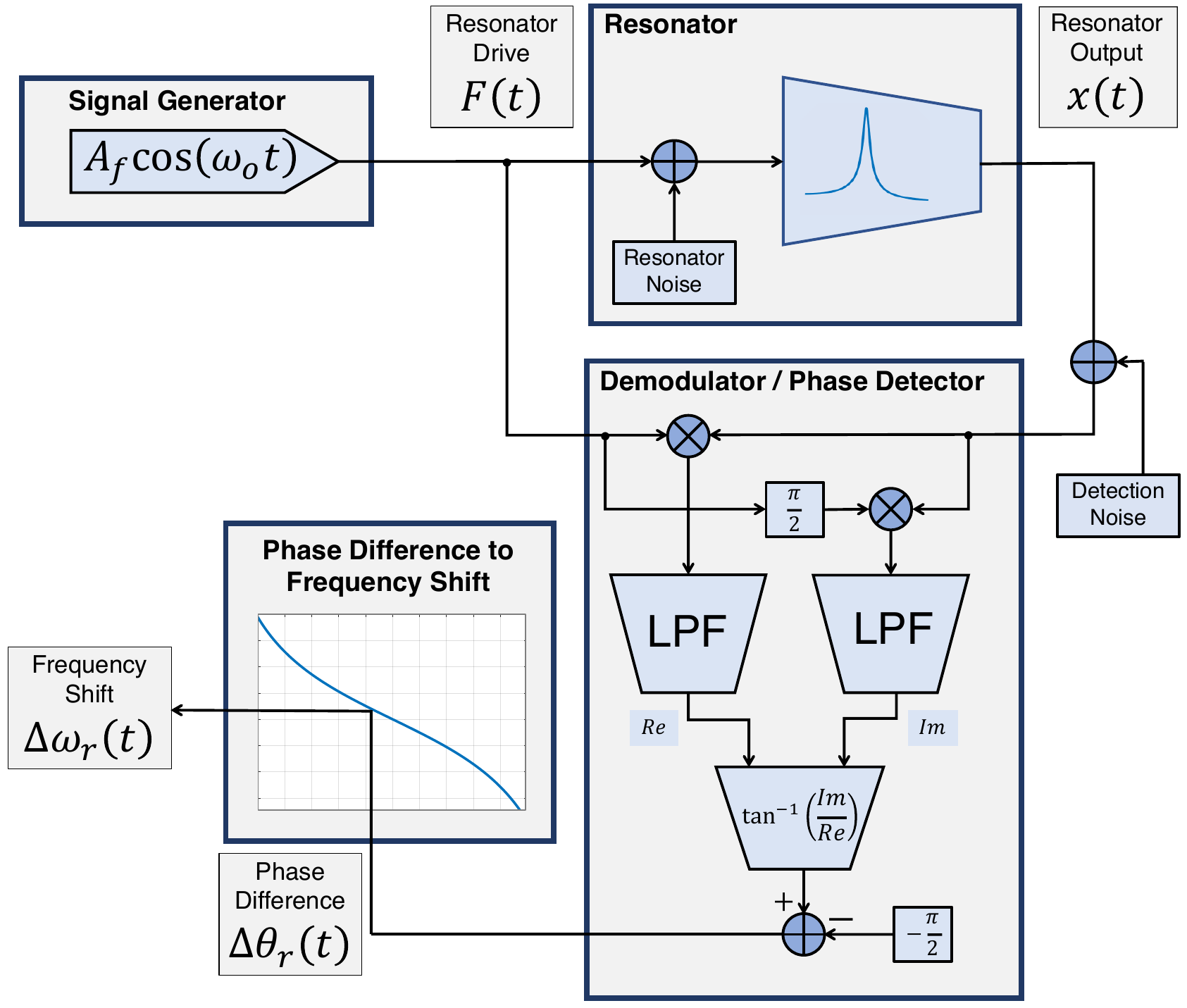}
\caption{Feedback-free (FF) scheme (figure based on~\cite[Fig.~2]{demir2019fundamental})}
\label{fig:openloop}
\end{figure}

In the feedback-free scheme, shown in Fig.~\ref{fig:openloop}, the
signal that drives the resonator has a fixed frequency. At
steady-state, the drive signal and the resonator output are at the
same (drive) frequency, but there is a (constant) phase difference between
them, equal to $-\tfrac{\pi}{2}$ if the drive frequency is equal to the
resonance frequency. Any resonance frequency shift is detected and
inferred by continually monitoring this phase difference~\cite{sadeghi2020frequency}, 
and mapping it to a frequency shift based on the (inverse) of the phase response of the
resonator given by
\begin{equation}
  \theta_{\textsf{\tiny r}}\pqty{t} =
  {{-\tfrac{\pi}{2}-\atan\pqty{\tau_{\textsf{\tiny r}}\,\Delta\omega_{\textsf{\tiny r}}\pqty{t}}}}\approx -\tfrac{\pi}{2}-\tau_{\textsf{\tiny r}}\,\Delta\omega_{\textsf{\tiny r}}\pqty{t}  
\label{eqn:phresp}
\end{equation}
As above, the phase response can be approximated as a linear function
around the resonance frequency. Thus, the map from the phase
difference to the frequency shift takes the following simple form
\begin{equation}
  \Delta\omega_{\textsf{\tiny r}}\pqty{t} = -\frac{\Delta\theta_{\textsf{\tiny r}}\pqty{t}}{\tau_{\textsf{\tiny r}}}
\label{eqn:phdifftofreqshift}
\end{equation}
We recall that the phase/frequency shift response above exhibits a
transient following a resonance frequency shift causing
event, as in \eqref{eqn:resonatormodelfrshiftforphasediffsimpultsoln},
arising from the inherent dynamics of the resonator. That is, the {\em
  full} impact of the event can not be instantaneously observed in the
phase response of the resonator. This observation is in contrast with
an apparent misconception in the nanomechanical sensors field that the
phase/frequency response of a resonator has an instantaneous nature,
whereas the amplitude response exhibits a transient dictated by its
inherent dynamics. In fact, both the amplitude and
phase exhibit inherent transient characteristics, as shown in
\eqref{eqn:drivenresonatortraresponsesoln} and
\eqref{eqn:resonatormodelfrshiftforphasediffsimpultsoln}.
Eqn.~\eqref{eqn:resonatormodelfrshiftforphasediffsimpultsoln} 
was verified experimentally with the measurement results 
reported in~\cite[Apdx~A]{sadeghi2020frequency}.   
We will delve deeper into this issue later in our treatment, when we compare
the feedback-free approach with the other closed-loop schemes.

The phase difference between two narrowband (nearly sinusoidal)
signals can be extracted using a narrowband {\sc IQ} demodulator as
shown in Fig.~\ref{fig:openloop}. Alternatively, one can use a Hilbert
transform based
scheme~\cite{benedetto1999principles,
  giridharagopal2012submicrosecond}.
Since the two schemes are equivalent for narrowband signals, the {\sc IQ}
demodulator is preferred due to its less complex nature when compared with the Hilbert
transform.

The {\sc IQ} demodulator uses low-pass filters ({\sc LPF}s) in order
to filter out the frequency components, that are at twice the
resonance frequency, produced by the multipliers. The bandwidths of
these filters need to be chosen to be less than (twice) the resonance
frequency. This imposes a further limitation on the bandwidth (speed)
of the feedback-free resonator tracking scheme, in addition to the
inherent low-pass filtering represented by \eqref{eqn:restranfun} due
to the resonator dynamics. For high quality factor resonators, the
bandwidth limitation imposed by the resonator characteristics is the
limiting factor.

The bandwidth of the {\sc LPF}s in the demodulator and inherent
resonator dynamics not only influence the transient response of the
feedback-free tracking scheme to frequency shifts caused by events of
interest, but also determine the noise performance and hence the
sensitivity together with the sources of noise. 

\subsection{Frequency-locked loop (FLL) based resonator tracking}
\label{sec:fll}

\begin{figure}
\centering
\includegraphics[width=1.0\columnwidth]{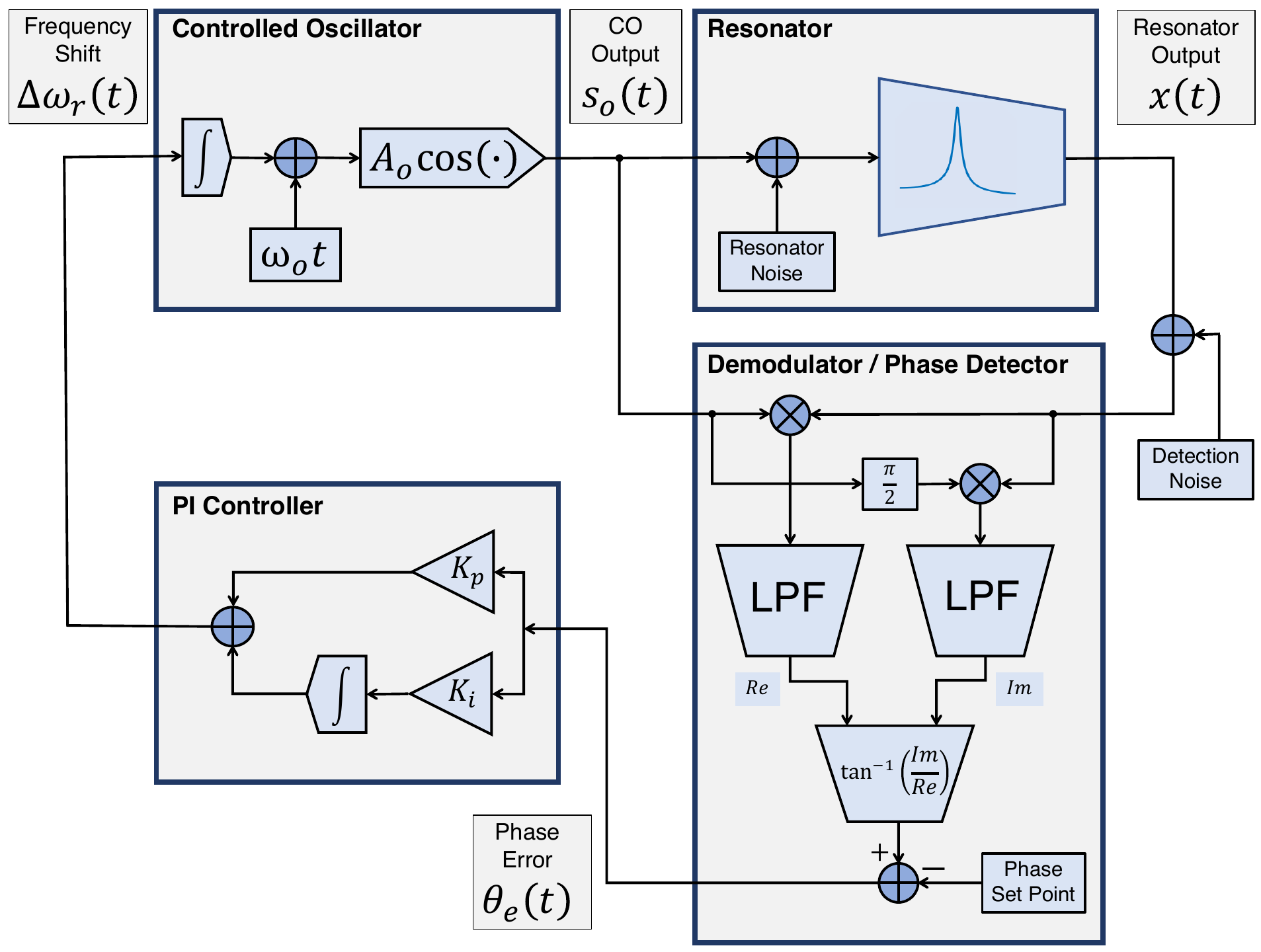}
\caption{Frequency-locked loop (FLL) (figure based on~\cite[Fig.~2]{demir2019fundamental})}
\label{fig:fll}
\end{figure}

In the closed-loop {\sc FLL} scheme, shown in Fig.~\ref{fig:fll}, the
frequency of the drive signal is continually updated with a
negative-feedback loop in such a way so that the resonator is always
operating at its resonance. This is accomplished by feeding the phase
difference (minus the phase set point $-\tfrac{\pi}{2}$) between the
drive signal and the resonator response as an error signal into a {\sc
  PI} controller. The output of the controller is used to set the
frequency of the signal generator, \ie a controlled oscillator ({\sc
  CO}). Thus, any resonance frequency shift will result in a nonzero
phase error, and a subsequent update in the controller output, which
can be directly used to continually monitor the frequency of
the {\sc CO} and hence the resonance frequency. In the {\sc FLL}
scheme, there is no need for an explicit map from phase error to
frequency shift. This is inherently accomplished by the {\sc FLL}
dynamics, without relying on information regarding the phase response
characteristics of the resonator.

The phase error signal in the {\sc FLL} scheme is produced in exactly
the same manner as the phase difference in the feedback-free aproach.
Thus, one may conclude that the speed of the {\sc FLL} response to
events of interest will be also limited by the inherent resonator
dynamics. However, the phase difference signals are used in a
different manner in the two schemes. In the feedback-free scheme, the
phase difference is directly mapped to the frequency shift, whereas in
the {\sc FLL}, it is further processed by the controller and also
shaped by the closed-loop dynamics. The response characteristics of
the {\sc FLL} may be tailored so that it is much faster, in fact, not
limited by the resonator dynamics. We present a precise, quantitative
analysis of the {\sc FLL} later in our treatment. Qualitatively, the
controller in the {\sc FLL} can be designed so that the
negative-feedback either speeds up or, if necessary, slows down the
{\sc FLL} with respect to the inherent resonator response, albeit with
implications for noise filtering bandwidth. This allows one to trade speed 
for accuracy or vice versa.  

\subsection{Self-sustained oscillator (SSO) based resonator tracking}
\label{sec:selfoscillating}

\begin{figure}
\centering
\includegraphics[width=1.0\columnwidth]{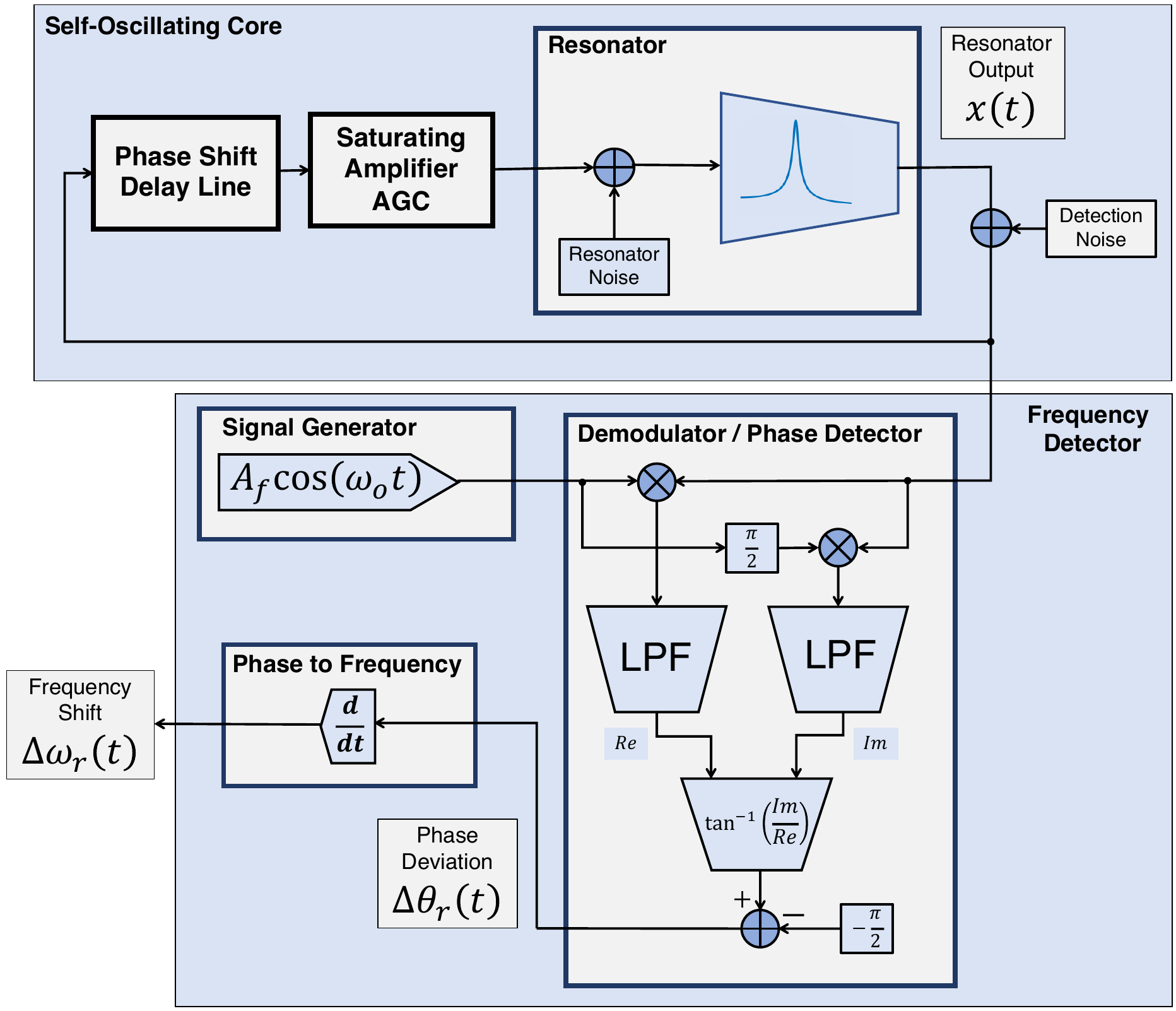}
\caption{Self-sustained oscillator (SSO) (figure based on~\cite[Fig.~2]{demir2019fundamental})}
\label{fig:selfoscillating}
\end{figure}

One embodiment of the self-sustained oscillator based resonator
tracking scheme is shown in Fig.~\ref{fig:selfoscillating}, composed
of a self-oscillating core based on positive-feedback and a separate
frequency detector. In Fig.~\ref{fig:selfoscillating}, the frequency
detector is realized using an {\sc IQ} demodulator that includes a
signal generator at a fixed frequency, which basically measures the
total phase difference between the nearly sinusoidal signal generated
by the self-oscillating core, called the oscillator output from now
on, and the sinusoidal signal output of the signal generator. If the
oscillator frequency is exactly equal to the signal generator
frequency, then the phase difference is constant. If the two
frequencies are different, then the phase difference is a linear
function of time with the slope as the frequency difference. Thus, in
order to compute the frequency difference, the phase difference is
differentiated {\em w.r.t.} to time as in
Fig.~\ref{fig:selfoscillating}. We note that the use of the
demodulator in Fig.~\ref{fig:selfoscillating} is quite different when
compared with its use, in Figs.~\ref{fig:openloop} and \ref{fig:fll},
in the feedback-free and the {\sc FLL} schemes, where the signal
generator that drives the resonator also drives one of the inputs of
the demodulator. On the other hand, in the frequency detector in
Fig.~\ref{fig:selfoscillating}, one demodulator input is set to the
self-generated oscillator output, and the other one is a fixed
frequency signal unrelated to the resonator drive or response. A
resonance frequency shift in the feedback-free scheme
results in a (steady-state) {\em time-invariant phase difference},
whereas for the self-sustained oscillator scheme, the resulting {\em
  phase difference changes linearly with time}. As a result, the
mapping of the phase difference to a frequency shift is done via
\eqref{eqn:phdifftofreqshift} in the feedback-free scheme, whereas
with a time derivative in the frequency detector in
Fig.~\ref{fig:selfoscillating}. There seems to be a confusion
regarding this issue in the literature. For instance, authors
in~\cite{giridharagopal2012submicrosecond} seem to be using a
feedback-free scheme, but employ a time derivative in order to map the
phase information into a frequency shift.

In Section~\ref{sec:selfoscstepresp}, we derived the (resonance)
frequency step response of the self-oscillating core, and obtained
\eqref{eqn:selfoscmodelphasestepresponse},
reproduced here for convenience:
\begin{equation}
  \theta_{\textsf{\tiny r}}\pqty{t} = \pqty{\omega_{\textsf{\tiny r2}}-\omega_{\textsf{\tiny r1}}}\:t + \theta_{\textsf{\tiny r1}} = \Delta\omega_{\textsf{\tiny e}}\:t + \theta_{\textsf{\tiny r1}} 
\nonumber
\end{equation}
The above phase response was derived for a sudden resonance frequency
shift of size $\Delta\omega_{\textsf{\tiny e}}$ that occurs at $t=0$. Thus, the
time-derivative of the phase deviation above produces an {\em
  instantaneous} frequency shift from 0 to $\Delta\omega_{\textsf{\tiny e}}$ at
$t=0$. The fact that a self-oscillating core produces an instantaneous
response to a resonance frequency shift was stated and exploited in
inventing the {\sc FM-AFM} scheme in~\cite{albrecht1991frequency},
however, a theoretical derivation was not provided. The theory and
analysis we have presented in Section~\ref{sec:selfoscstepresp} provides
the rigorous theoretical foundation for this observation. Our theory in
Section~\ref{sec:theory}, covering all three resonator tracking
schemes in a unified manner, clarifies the misconceptions and
confusions in the field regarding the transient and instantaneous
nature of amplitude/phase/frequency step responses of resonators in various
configurations.  We note that the phase step response in all of the schemes
is a {\em continuous function of time}, does not exhibit any
unphysical jumps. However, the {\em continuous phase response} maps 
to a {\em continuous frequency step response} in the
feedback-free case,
whereas to a {\em discontinuous frequency step response} in the
self-oscillating scheme. Even though the inherent frequency step response
of a self-oscillating core is indeed instantaneous, any practical
frequency detection scheme will come with a bandwidth and speed
limitation in measuring this instantaneous frequency 
response~\cite{albrecht1991frequency,mascaro2019review}. For
instance, the frequency detection scheme in
Fig.~\ref{fig:selfoscillating} based on an {\sc IQ} demodulator uses
{\sc LPF}s with a bandwidth less than (twice) the resonance
frequency. Thus, the frequency shift output of the frequency detector
will be smoothed with a time-constant that is dictated by the
bandwidths of the {\sc LPF}s, even though the inherent frequency
step response of the self-oscillating core is sharp. Other frequency
detection schemes also have bandwidth limitations. For instance, a
{\sc PLL} which locks the phase and frequency of a signal generator to
the oscillator output has a certain loop bandwidth that dictates its
response speed.

\section{Comparative characterization of resonator tracking schemes}
\label{sec:analysis}

We next perform a comparative analysis of the three resonator tracking
schemes under consideration, using the theory developed in
Section~\ref{sec:theory}. There are two performance criteria we
consider, simply put, {\em speed} and {\em accuracy}. We quantify
speed in terms of how quickly the system responds to resonance
frequency jumps, \ie with {\em Frequency Step Response} ({\sc
  FSTR}). Accuracy, or sensitivity, is assessed based on various 
characterizations of the frequency fluctuations due to noise in the
frequency shift output of the system in tracking mode.

\begin{figure}
\centering
\includegraphics[width=1.0\columnwidth]{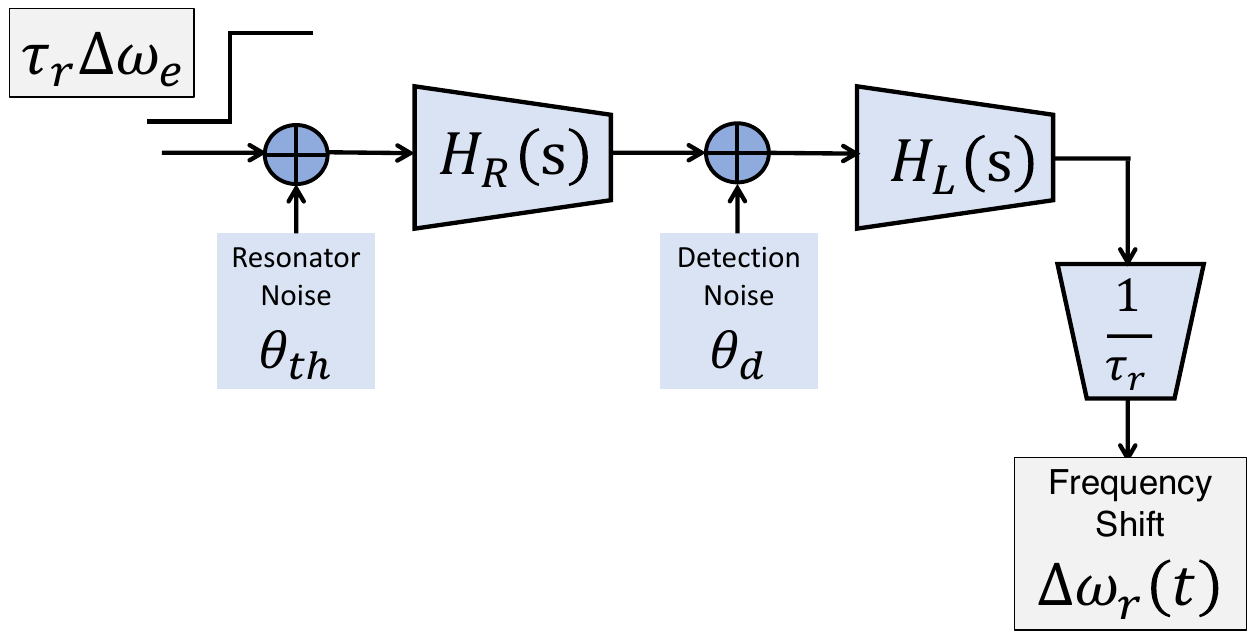}
\caption{Phase-domain model for {\sc FF} scheme}
\label{fig:openloopphase}
\end{figure}

\begin{figure}
\centering
\includegraphics[width=1.0\columnwidth]{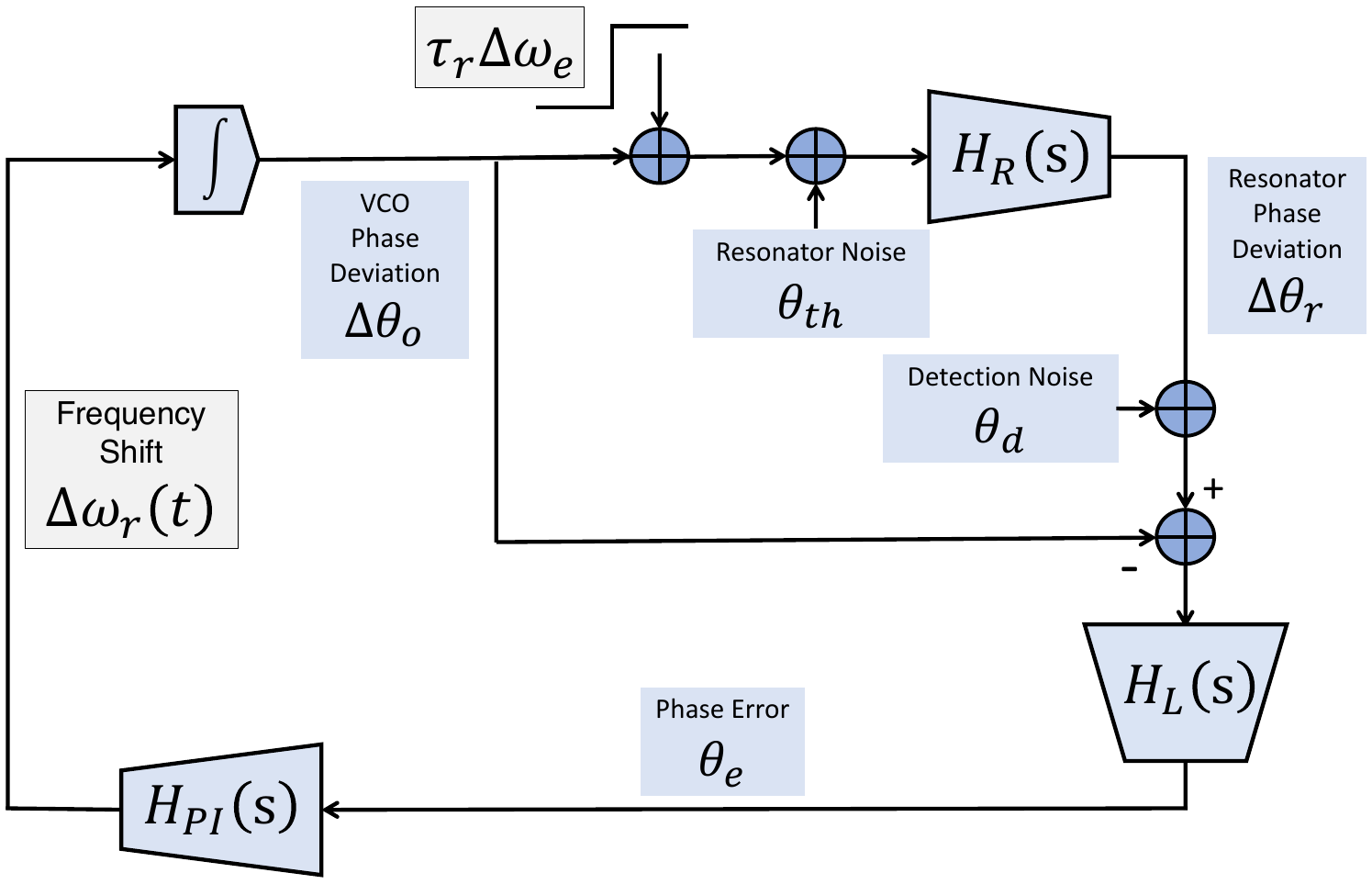}
\caption{Phase-domain model for {\sc FLL} (figure based on~\cite[Fig.~3]{demir2019fundamental})}
\label{fig:fllphase}
\end{figure}

\begin{figure}
\centering
\includegraphics[width=1.0\columnwidth]{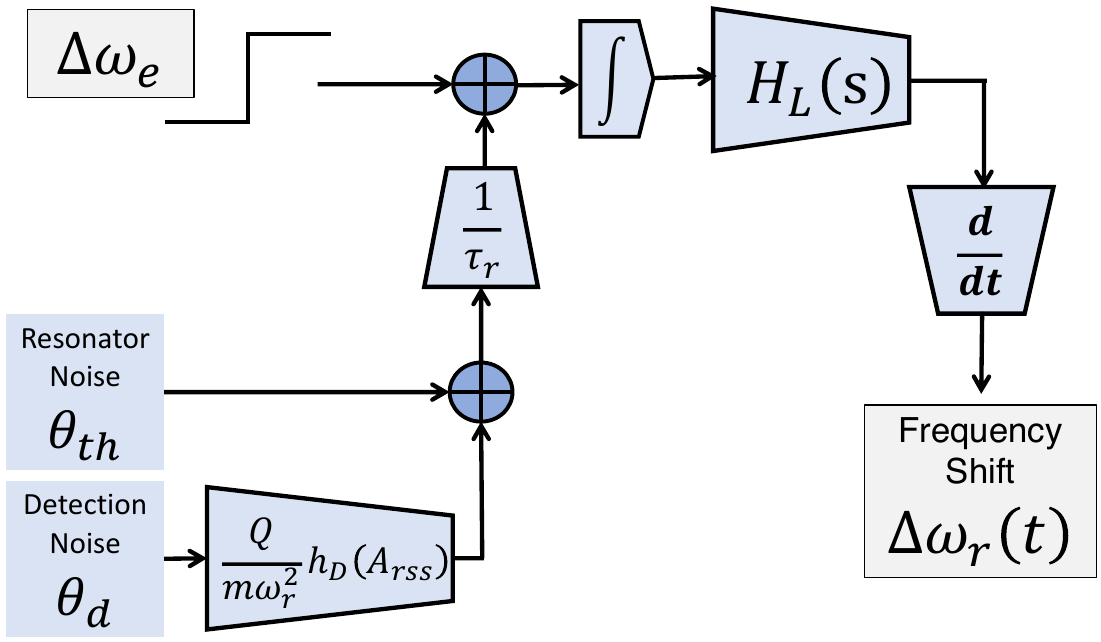}
\caption{Phase-domain model for {\sc SSO}}
\label{fig:selfoscillatingphase}
\end{figure}

In order to simplify performance characterization and enable
comparative analysis, we distill and capture the results of
the theory developed in Section~\ref{sec:theory} by creating simplified
phase/frequency domain models for all of the tracking schemes, shown as
block diagrams in Figs.~\ref{fig:openloopphase},~\ref{fig:fllphase}
and~\ref{fig:selfoscillatingphase}. In these diagrams,
$\Delta\omega_{\textsf{\tiny e}}$ represents a resonance frequency step change due to
an event of interest. All functions of $\las$ represent Laplace domain
transfer functions. $H_\textsf{\tiny R}\pqty{\las}$, given in \eqref{eqn:restranfun},
models the inherent resonator dynamics. $H_\textsf{\tiny L}\pqty{\las}$ represents the
low-pass filtering (with a bandwidth less then (twice) the resonance
frequency and a unity gain for DC inputs) in the {\sc IQ} demodulator. 
$H_{\textsf{\tiny PI}}\pqty{s}$ is for the controller
\begin{equation}
H_{\textsf{\tiny PI}}\pqty{\las} = K_\textsf{\tiny p} + \frac{K_\textsf{\tiny i}}{\las}.
\label{eqn:picontrollertransfun}
\end{equation} 
We consider a {\it proportional-integral} ({\sc PI}) controller in this work, as it is very commonly used. 
Our treatment in the paper, however, can be applied to other controllers in a straightforward manner. 
The (white) noise sources representing the thermomechanical noise of
the resonator, $\theta_{\textsf{\tiny th}}$, and the detection noise, $\theta_{\textsf{\tiny d}}$, are
characterized as given in \eqref{eqn:thetan} and
\eqref{eqn:detectionnoiseinoutput}.

\subsection{Frequency step response: Speed}
\label{sec:speed}

Frequency step response can be easily characterized by computing the
transfer function from the frequency step input $\Delta\omega_{\textsf{\tiny e}}$ to
the frequency shift output $\Delta\omega_{\textsf{\tiny r}}$, using the block diagrams
in Figs.~\ref{fig:openloopphase},~\ref{fig:fllphase}
and~\ref{fig:selfoscillatingphase}:
\begin{equation}
  \begin{aligned}
    H^{\textsf{\tiny FF}}_{\textsf{\tiny FST}}\pqty{\las} & = H_\textsf{\tiny R}\pqty{\las}\:H_\textsf{\tiny L}\pqty{\las}\\
    H^{\textsf{\tiny FLL}}_{\textsf{\tiny FST}}\pqty{\las} & = \frac{\pqty{\las K_\textsf{\tiny p}+ K_\textsf{\tiny i}} H_\textsf{\tiny L}\pqty{\las}}{\las^2+\frac{\las}{\tau_{\textsf{\tiny r}}}+\pqty{\las K_\textsf{\tiny p}+ K_\textsf{\tiny i}} H_\textsf{\tiny L}\pqty{\las}}\\
    H^{\textsf{\tiny SSO}}_{\textsf{\tiny FST}}\pqty{\las} & = H_\textsf{\tiny L}\pqty{\las}
  \end{aligned}
  \label{eqn:tranfunfstr}
\end{equation}
The step response is then simply given by
\begin{equation}
  f_{\textsf{\tiny str}}\pqty{t} = {\cal
    L}^{-1}\Bqty{\frac{H_{\textsf{\tiny FST}}\pqty{\las}}{\las}} 
\label{eqn:fstr}
\end{equation}
where ${\cal L}^{-1}\Bqty{\bullet}$ denotes the inverse Laplace transform, with
  $f_{\textsf{\tiny str}}\pqty{0} = 0$ and $f_{\textsf{\tiny str}}\pqty{t\rightarrow \infty} = 1$
  since $H_{\textsf{\tiny FST}}\pqty{\las\rightarrow 0}=1$.  

  In \eqref{eqn:tranfunfstr}, $H^{\textsf{\tiny FF}}_{\textsf{\tiny FST}}\pqty{\las}$ and
  $H^{\textsf{\tiny SSO}}_{\textsf{\tiny FST}}\pqty{\las}$ are simple, but
  $H^{\textsf{\tiny FLL}}_{\textsf{\tiny FST}}\pqty{\las}$ needs some clarification. $H^{\textsf{\tiny FLL}}_{\textsf{\tiny FST}}\pqty{\las}$ was obtained by writing
  and solving an equation~\cite{demir2019fundamental} based on
  the block diagram in Fig.~\ref{fig:fllphase}. 
  We choose $K_\textsf{\tiny p}$ and $K_\textsf{\tiny i}$, as suggested
in~\cite{olcum2015high,demir2019fundamental},
\begin{equation}
K_\textsf{\tiny p} = \omega_{\textsf{\tiny FLL}} \quad,\quad K_\textsf{\tiny i} = \frac{\omega_{\textsf{\tiny FLL}} }{\tau_{\textsf{\tiny r}}}  
\label{eqn:PIpars}  
\end{equation}
where $\omega_{\textsf{\tiny FLL}}$ is the desired {\sc FLL} loop
bandwidth. The above choice yields~\cite{olcum2015high,demir2019fundamental}  
\begin{equation}
\begin{aligned}
H^{\textsf{\tiny FLL}}_{\textsf{\tiny FST}}\pqty{\las}  = 
\frac{H_\textsf{\tiny L}\pqty{\las}}{H_\textsf{\tiny L}\pqty{\las} + \frac{\las}{\omega_{\textsf{\tiny FLL}} }}
\end{aligned}
\label{eqn:looptrfunFLL}
\end{equation}
The simplified transfer function above, with a low-pass
characteristics, is independent of, and not limited by, the resonator
dynamics, as a direct consequence of the choice in
\eqref{eqn:PIpars}. Thus, while the open-loop {\sc FF} approach
suffers from a bandwidth limitation imposed by the inherent resonator
dynamics, both the {\sc FLL} and {\sc SSO} schemes do not.  This is
enabled by feedback (negative-feedback for {\sc FLL}, and
positive-feedback for {\sc SSO}), which in a sense speeds up the
response of the system, albeit at the expense of worse noise
performance, as we discuss later. In the case of {\sc FLL}, this
happens through a pole-zero
cancellation~\cite{olcum2015high,demir2019fundamental} in the transfer
function.

For a detailed characterization of speed, one can simply evaluate
\eqref{eqn:tranfunfstr} and \eqref{eqn:fstr} and compute the settling
time of {\sc FSTR}. Roughly, we can conclude that the response time
constant of both the {\sc FLL} and the {\sc SSO} schemes can be as
small as $\tfrac{1}{\omega_{\textsf{\tiny r}}}$ limited by the required low-pass
filtering in the demodulator/phase detector, whereas for the {\sc FF}
scheme, it is at least $\tfrac{Q}{\omega_{\textsf{\tiny r}}}$ limited by the inherent
resonator dynamics.  One can, of course, for all schemes, make the
response slower in order to obtain better steady-state tracking noise
performance, \ie for more accuracy, as we discuss below.

\subsection{Spectral density of frequency fluctuations: Accuracy}
\label{sec:accuracy}

In order to characterize the fluctuations due to noise in the
frequency shift output of the system, we need to compute transfer
functions from the two noise sources $\theta_{\textsf{\tiny th}}$ and $\theta_{\textsf{\tiny d}}$ to the
output $\Delta\omega_{\textsf{\tiny r}}$. This can be done easily using the block diagrams
in Figs.~\ref{fig:openloopphase},~\ref{fig:fllphase}
and~\ref{fig:selfoscillatingphase}:
\begin{equation}
  \begin{aligned}
    H^{\textsf{\tiny FF}}_{\theta_{\textsf{\tiny th}}}\pqty{\las} & =
    \frac{1}{\tau_{\textsf{\tiny r}}}\:H_\textsf{\tiny R}\pqty{\las}\:H_\textsf{\tiny L}\pqty{\las}\\
        H^{\textsf{\tiny FF}}_{\theta_{\textsf{\tiny d}}}\pqty{\las} & =
        \frac{1}{\tau_{\textsf{\tiny r}}}\:H_\textsf{\tiny L}\pqty{\las} \\
    H^{\textsf{\tiny FLL}}_{\theta_{\textsf{\tiny th}}}\pqty{\las} & = \frac{1}{\tau_{\textsf{\tiny r}}}\frac{\pqty{\las
        K_\textsf{\tiny p}+ K_\textsf{\tiny i}}
      H_\textsf{\tiny L}\pqty{\las}}{\las^2+\frac{\las}{\tau_{\textsf{\tiny r}}}+\pqty{\las K_\textsf{\tiny p}+ K_\textsf{\tiny i}}
      H_\textsf{\tiny L}\pqty{\las}}\\
    H^{\textsf{\tiny FLL}}_{\theta_{\textsf{\tiny d}}}\pqty{\las} & = \frac{1}{\tau_{\textsf{\tiny r}}}\frac{1}{H_\textsf{\tiny R}\pqty{\las}}\frac{\pqty{\las
        K_\textsf{\tiny p}+ K_\textsf{\tiny i}}
      H_\textsf{\tiny L}\pqty{\las}}{\las^2+\frac{\las}{\tau_{\textsf{\tiny r}}}+\pqty{\las K_\textsf{\tiny p}+ K_\textsf{\tiny i}}
     H_\textsf{\tiny L}\pqty{\las}}\\   
    H^{\textsf{\tiny SSO}}_{\theta_{\textsf{\tiny th}}}\pqty{\las} & =
    \frac{1}{\tau_{\textsf{\tiny r}}}\:H_\textsf{\tiny L}\pqty{\las}\\
        H^{\textsf{\tiny SSO}}_{\theta_{\textsf{\tiny d}}}\pqty{\las} & = \frac{Q}{m\,\omega^2_{r}}h_\textsf{\tiny D}\pqty{A_{\textsf{\tiny rss}}}\:\frac{1}{\tau_{\textsf{\tiny r}}}\:H_\textsf{\tiny L}\pqty{\las}
  \end{aligned}
  \label{eqn:tranfunthetand}
\end{equation}
The transfer functions above are used in computing the spectral
density of the fractional frequency fluctuations, \ie
$y_{\textsf{\tiny r}}=\tfrac{\Delta\omega_{\textsf{\tiny r}}}{\omega_{\textsf{\tiny r}}}$, as follows
\begin{equation}
  \begin{aligned}
    \cS_{y_{\textsf{\tiny r}}}\pqty{\omega} = \frac{\vqty{H_{\theta_{\textsf{\tiny th}}}\pqty{j\omega}}^2\:\cS_{\theta_{\textsf{\tiny th}}}\pqty{\omega} + 
      \vqty{H_{\theta_{\textsf{\tiny d}}}\pqty{j\omega}}^2\:\cS_{\theta_{\textsf{\tiny d}}}\pqty{\omega}}{\omega_{\textsf{\tiny r}}^2}
  \end{aligned}
  \label{eqn:specy}
\end{equation}
where $\cS_{\theta_{\textsf{\tiny th}}}\pqty{j\omega}$ and
$\cS_{\theta_{\textsf{\tiny d}}}\pqty{j\omega}$ are the (white) spectral densities of 
$\theta_{\textsf{\tiny th}}$ and $\theta_{\textsf{\tiny d}}$, given by
\begin{equation}
  \begin{aligned}
    \cS_{\theta_{\textsf{\tiny th}}}\pqty{\omega} &=
      \bqty{\frac{1}{A_{\textsf{\tiny rss}}}\frac{Q}{m\,\omega^2_{r}}}^2\:4\,m\,\frac{\omega_{\textsf{\tiny r}}}{Q}\,k_{\textsf{\tiny B}}\,T
      = \frac{4\,Q\,k_{\textsf{\tiny B}}\,T}{m\,\omega_{\textsf{\tiny r}}^3\,A_{\textsf{\tiny rss}}^2}\\
    \cS_{\theta_{\textsf{\tiny d}}}\pqty{\omega} &= {\cal K}_{\textsf{\tiny d}}^2\:\cS_{\theta_{\textsf{\tiny th}}}\pqty{\omega}
  \end{aligned}
  \label{eqn:specthetand}
\end{equation}
The transfer functions in \eqref{eqn:tranfunthetand} all have 
low-pass characteristics. The spectral density of fractional frequency
fluctuations, $\cS_{y_{\textsf{\tiny r}}}\pqty{\omega}$, is thus in the form of low-pass
filtered white noise. The value of the spectral density at zero (low)
frequency, $\cS_{y_{\textsf{\tiny r}}}\pqty{\omega=0}$, is of interest, since it determines
the accuracy of the resonator tracking scheme at steady-state, after
the step response has settled following a resonance frequency shift
due to an event of interest. $\cS_{y_{\textsf{\tiny r}}}\pqty{\omega=0}$ is a measure of
accuracy/sensitivity that is independent of speed/bandwidth, as
opposed to, for instance, mean-squared fluctuations that can be computed by
integrating $\cS_{y_{\textsf{\tiny r}}}\pqty{\omega}$ over the system bandwidth.
While the bandwidth/speed of a flexible resonator tracking scheme may
be adjusted for speed versus accuracy trade-off,
$\cS_{y_{\textsf{\tiny r}}}\pqty{\omega=0}$ quantifies a fundamental sensitivity
limitation. We evaluate $\cS_{y_{\textsf{\tiny r}}}\pqty{0}$ using
\eqref{eqn:tranfunthetand}, \eqref{eqn:specy} and
\eqref{eqn:specthetand} to obtain
  \begin{align}
    \nonumber
    \cS_{y_{\textsf{\tiny r}}}^{\textsf{\tiny FF}}\pqty{0} & =
    \frac{k_{\textsf{\tiny B}}\,T}{m\,Q\,\omega_{\textsf{\tiny r}}^3\,A_{\textsf{\tiny rss}}^2}\,\pqty{1+{\cal
        K}_{\textsf{\tiny d}}^2} \\
  \label{eqn:specyatzero}
    \cS_{y_{\textsf{\tiny r}}}^{\textsf{\tiny FLL}}\pqty{0} & =
    \frac{k_{\textsf{\tiny B}}\,T}{m\,Q\,\omega_{\textsf{\tiny r}}^3\,A_{\textsf{\tiny rss}}^2}\,\pqty{1+{\cal
        K}_{\textsf{\tiny d}}^2}\\
    \cS_{y_{\textsf{\tiny r}}}^{\textsf{\tiny SSO}}\pqty{0} & =
    \frac{k_{\textsf{\tiny B}}\,T}{m\,Q\,\omega_{\textsf{\tiny r}}^3\,A_{\textsf{\tiny rss}}^2}\,\bqty{1+
      \pqty{{\cal
          K}_{\textsf{\tiny d}}\,\frac{Q}{m\,\omega^2_{r}}h_\textsf{\tiny D}\pqty{A_{\textsf{\tiny rss}}} }^2}
    \nonumber
  \end{align}  
We observe that the {\sc FF} and {\sc FLL} schemes
have the same performance, whereas for {\sc SSO}, the performance
is degraded due to the amplification of detection noise through the
amplifier in the positive feedback loop as well as the resonator
itself. The {\sc FF} and {\sc FLL} schemes do not suffer from
amplified detection noise, since the resonator is directly driven with
a clean signal generator instead of an amplified version of the signal
transduced from its own output. If thermomechanical noise is very well
resolved, that is, if detection noise is negligible in comparison, with
${\cal K}_{\textsf{\tiny d}}\ll 1$, then all three schemes have the same performance as
characterized by \eqref{eqn:specyatzero}.

\subsection{Speed versus Accuracy}
\label{sec:speedvsaccuracy}

There is a direct trade-off between accuracy and speed in resonator
tracking systems. Any mechanism that speeds up the response is bound
to be more susceptible to noise. Conversely, more noise filtering for improved accuracy, in some form or another, will result in a slower response. The accuracy versus speed trade-off characteristics 
of resonator tracking schemes are not always well
understood and articulated in the literature. We next
compare the three schemes under consideration with respect to their
speed-accuracy trade-off characteristics. We first consider the case where thermomechanical noise is well resolved, \ie detection noise is negligible in comparison. We then consider detection noise. 

For high-$Q$ resonators, the response speed of the {\sc FF} scheme is determined by the inherent resonator dynamics, as 
discussed in Section~\ref{sec:speed}. The response time constant is equal to the resonator time constant $\tfrac{2 Q}
{\omega_{\textsf{\tiny r}}}$. In accordance, the thermomechanical noise of 
the resonator is filtered by the resonator transfer function, with a bandwidth of $\tfrac{\omega_{\textsf{\tiny r}}}{2 Q}$ as in \eqref{eqn:tranfunthetand}. As a result, when $Q$ is increased, thermomechanical noise bandwidth decreases in proportion, and at the same time, so does the noise spectral density within that bandwidth, as given in \eqref{eqn:specyatzero}. Thus, one obtains a {\em quadratic reduction} in the noise power that is passed through. However, this comes at the expense of increased response time that is proportional to $Q$. Thus, in the {\sc FF} scheme, increasing $Q$ results in a quadratic improvement in accuracy at the expense of a linear degradation in speed,
assuming that thermomechanical noise is dominant. The bandwidth of the low-pass filter $H_\textsf{\tiny L}\pqty{s}$ in the {\sc IQ} demodulator is typically larger than $\tfrac{\omega_{\textsf{\tiny r}}}{2 Q}$, with a maximum value of $\omega_{\textsf{\tiny r}}$. One may set the bandwidth of $H_\textsf{\tiny L}\pqty{s}$ to a value smaller than $\tfrac{\omega_{\textsf{\tiny r}}}{2 Q}$. This would result in further filtering for the thermomechanical noise of the resonator, as in \eqref{eqn:tranfunthetand}, in addition to the one provided by the inherent resonator dynamics, but at the same time further slowing down of the response.  

The {\sc FLL} scheme features a flexible speed-accuracy trade-off characteristics. The flexibility is brought about by feedback and the adjustable controller parameters. The transfer function $H^{\textsf{\tiny FLL}}_{\textsf{\tiny FST}}\pqty{\las}$ in \eqref{eqn:tranfunfstr} that dictates the frequency step response and $H^{\textsf{\tiny FLL}}_{\theta_{\textsf{\tiny th}}}\pqty{\las}$ in \eqref{eqn:tranfunthetand} that filters thermomechanical noise are the same. The bandwidth and characteristics of this filter that has a low-pass characteristics may be adjusted via the controller parameters $K_\textsf{\tiny p}$ and $K_{\textsf{\tiny i}}$, which were chosen in \eqref{eqn:PIpars} so that the prescribed bandwidth is independent of the resonator characteristics, in particular, its $Q$. As a result, when $Q$ is increased, the {\sc FLL} scheme does not suffer from a speed degradation. At the same time, thermomechanical noise spectral density decreases since it is inversely proportional to $Q$. 
When compared with the {\sc FF} scheme, the flexibility of the {\sc FLL} scheme enables one to give up the quadratic improvement in accuracy for only a linear improvement but with no degradation in speed. 
One can also design the controller in such a way so that one achieves better than linear improvement in accuracy with some degradation in speed. In fact, one may choose to endure worse than linear degradation in speed for a better than quadratic improvement in accuracy. In this case, the bandwidth of the {\sc FLL} would be designed to be even smaller than $\tfrac{\omega_{\textsf{\tiny r}}}{2 Q}$ that is imposed by the inherent resonator characteristics. The bandwidth adjustment for an {\sc FLL} can be done in a manner that is independent of the resonator characteristics. 

For the {\sc SSO} scheme, the transfer function $H^{\textsf{\tiny SSO}}_{\textsf{\tiny FST}}\pqty{\las}$ in \eqref{eqn:tranfunfstr} that dictates the frequency step response and $H^{\textsf{\tiny SSO}}_{\theta_{\textsf{\tiny th}}}\pqty{\las}$ in \eqref{eqn:tranfunthetand} that filters thermomechanical noise are the same, and determined by the low-pass filter $H_\textsf{\tiny L}\pqty{s}$ in the {\sc IQ} demodulator (frequency detector).  
The {\sc SSO} scheme also features a flexible speed-accuracy characteristics. The bandwidth of  $H_\textsf{\tiny L}\pqty{s}$ may be adjusted, with a maximum value of $\omega_{\textsf{\tiny r}}$, with the possibility of being smaller than $\tfrac{\omega_{\textsf{\tiny r}}}{2 Q}$. While speed is proportional to this bandwidth, accuracy is inversely proportional. Thus, the speed-accuracy trade-off characteristics of the {\sc SSO} scheme is similar to the one offered by the
{\sc FLL}, in the sense that bandwidth can be adjusted, independent of the resonator characteristics such as $Q$. This was the main motivation in the invention of the {\sc FM-AFM} technique in~\cite{albrecht1991frequency}, which is an {\sc SSO} based scheme. 
However, the {\sc FLL} scheme is more versatile. The loop transfer function that determines the frequency step response and thermomechanical noise filtering, given in \eqref{eqn:looptrfunFLL}, can be chosen to have a bandwidth that is less than the bandwidth of $H_\textsf{\tiny L}\pqty{s}$ in the demodulator. 

Detection noise in  the {\sc FF} scheme is simply filtered by the low-pass filter $H_\textsf{\tiny L}\pqty{s}$ in the {\sc IQ} demodulator. 
In the case of {\sc FLL}, nature of detection noise filtering is dependent on controller parameters. With controller parameters as in \eqref{eqn:PIpars}, detection noise transfer function of an {\sc FLL} takes the form 
\begin{equation}
H^{\textsf{\tiny FLL}}_{\theta_{\textsf{\tiny d}}}\pqty{\las} = \frac{1+\las\,\tau_{\textsf{\tiny r}}}{\tau_{\textsf{\tiny r}}}\:\frac{H_\textsf{\tiny L}\pqty{\las}}{H_\textsf{\tiny L}\pqty{\las} + \frac{\las}{\omega_{\textsf{\tiny FLL}} }}
\end{equation}
When $\omega_{\textsf{\tiny FLL}}\,\tau_{\textsf{\tiny r}} \gg 1$, \ie the loop bandwidth is larger than the resonator linewidth~\cite{sadeghi2020frequency}, this transfer function satisfies 
\begin{equation}
\begin{aligned}
H^{\textsf{\tiny FLL}}_{\theta_{\textsf{\tiny d}}}\pqty{0} & = \frac{1}{\tau_{\textsf{\tiny r}}} \\
H^{\textsf{\tiny FLL}}_{\theta_{\textsf{\tiny d}}}\pqty{j\omega} & \approx \omega_{\textsf{\tiny FLL}}\:H_\textsf{\tiny L}\pqty{j\omega}\quad\text{for}\quad\omega>\omega_{\textsf{\tiny FLL}} 
\end{aligned}
\end{equation}
Thus, the above implies that {\sc FLL} loop dynamics in fact amplifies (with respect to the {\sc  FF} scheme, see \eqref{eqn:tranfunthetand}) detection noise at frequencies above the loop bandwidth $\omega_{\textsf{\tiny FLL}}$, with an amplification factor of $\omega_{\textsf{\tiny FLL}}\,\tau_{\textsf{\tiny r}}=2 Q\,\tfrac{\omega_{\textsf{\tiny FLL}}}{\omega_{\textsf{\tiny r}}} = \tfrac{\omega_{\textsf{\tiny FLL}}}{\Gamma}$, where $\Gamma=\tfrac{\omega_{\textsf{\tiny r}}}{2 Q}$ is the (one-sided) linewidth of the resonator. The ramifications of this detection noise amplification predicted by our theory was experimentally observed in the measurements reported in~\cite[Fig.~5]{sadeghi2020frequency} for a closed-loop {\sc FLL} scheme. On the other hand, when the loop bandwidth $\omega_{\textsf{\tiny FLL}}$ is chosen to be {\em smaller} than the resonator linewidth $\Gamma$, detection noise is in fact {\em attenuated}  (with respect to the {\sc  FF} scheme). This prediction of our theory was also experimentally verified in~\cite[Fig.~5]{sadeghi2020frequency}.   

The detection noise amplification in the {\sc FLL} scheme can be alleviated to a degree by reducing the bandwidth of $H_\textsf{\tiny L}\pqty{s}$ down to $\omega_{\textsf{\tiny FLL}}$. Furthermore, noise amplification above the loop bandwidth has minimal impact on the accuracy performance of the {\sc FLL} scheme as we show later.   
On  the other hand, the {\sc SSO} scheme suffers from detection noise amplification all the way down to zero frequency, in the amount $\tfrac{Q}{m\,\omega^2_{r}}h_\textsf{\tiny D}\pqty{A_{\textsf{\tiny rss}}}$.  This will result in an accuracy performance degradation that can not be circumvented even when one is willing to trade-off speed performance for it.   

\subsection{Figure-of-Merit: Mean-Squared Error}
\label{sec:fstdev}

Following our somewhat qualitative treatment of speed versus accuracy trade-offs,  
we next perform a quantitative evaluation of the three sensor schemes and compare them in detail.
As a figure-of-merit that quantifies speed and accuracy in a unified manner, we use {\em Root Mean Squared Error} ($\text{\small \textsf{RMSE}}$). $\text{\small \textsf{RMSE}}$ combines {\it frequency step response} ({\sc
  FSTR}) for speed, with {\em Allan Deviation} ($\text{\small \textsf{AD}}$) for accuracy. 
{\sc FSTR} was defined in Section~\ref{sec:speed}.  

\begin{figure*}
\centering
\begin{minipage}{\textwidth}
\begin{mdframed}[style=fig,backgroundcolor=solarizedBase]
\centering
\includegraphics[width=\textwidth]{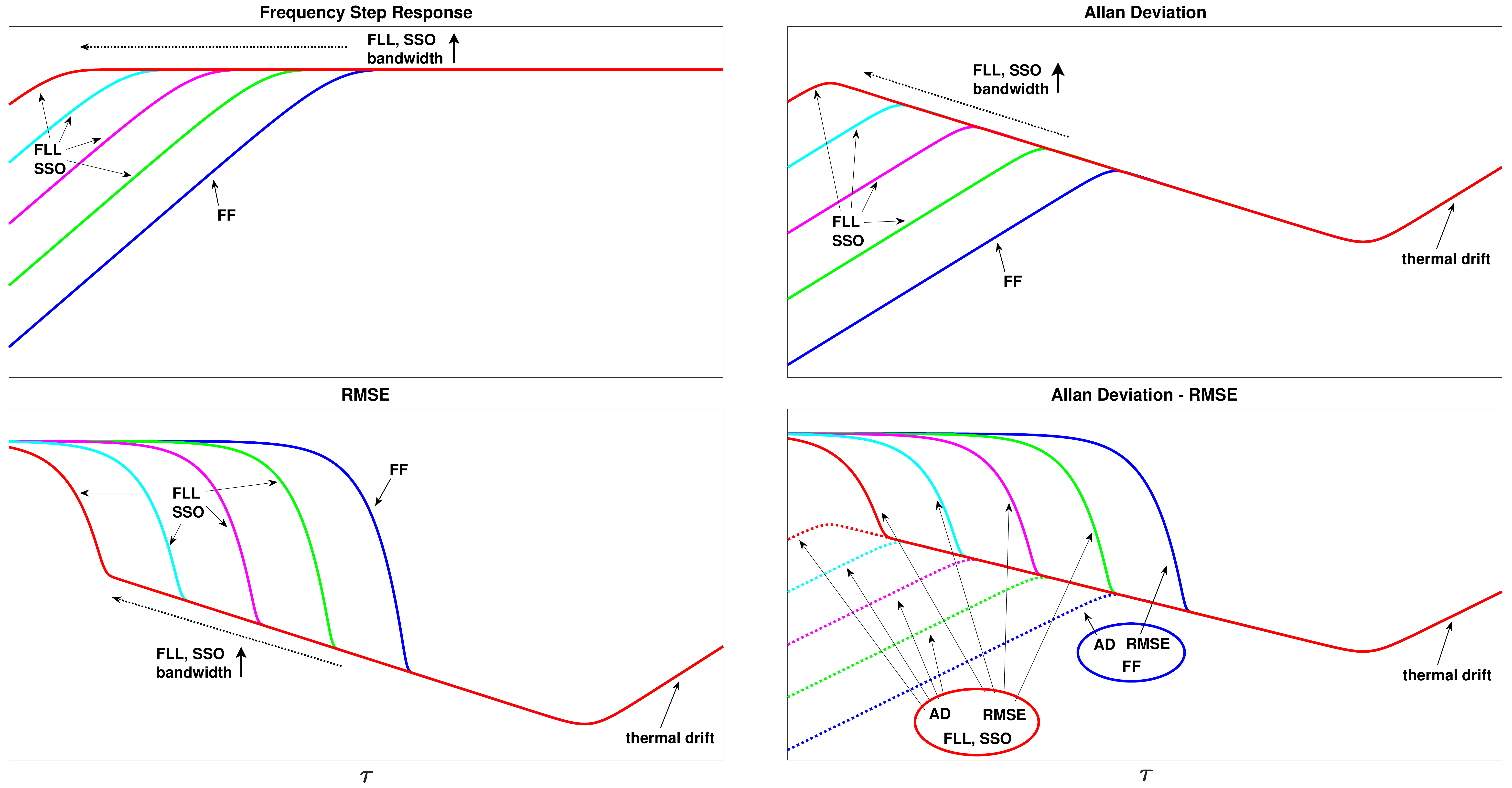}
\end{mdframed}
\end{minipage}
\caption{Frequency Step Response, Allan Deviation and Root Mean Squared Error}
\label{fig:fomstr}
\end{figure*}

$\text{\small \textsf{AD}}$~\cite{allan1966statistics,kroupa1983frequency,rubiola2009phase,makdissi2010,wiki:Allanvariance}
is the standard measure of frequency stability~\cite{IEEE1139}, established in the frequency standards community but also widely
used for resonant sensors. {\em Allan Variance} can be computed with 
\begin{equation}
\sigma_y^2\pqty{\tau} = \frac{4}{\pi\tau^2}\:\int_{-\infty}^{+\infty}
\frac{\bqty{\sin\pqty{\frac{\omega\,\tau}{2}}}^4}{\omega^2}\:\cS_{y_{\textsf{\tiny r}}}\pqty{\omega}\,\dd \omega  
\label{eqn:allanvarfrompsd}  
\end{equation}
where $\cS_{y_{\textsf{\tiny r}}}\pqty{\omega}$ is the spectrum of fractional frequency fluctuations given in \eqref{eqn:specy}, and $\tau$ is the {\em averaging time} for the fluctuations. 
$\text{\small \textsf{AD}}$ $\sigma_y\pqty{\tau}$ is uniquely determined by $\cS_{y_{\textsf{\tiny r}}}\pqty{\omega}$, as dictated by \eqref{eqn:allanvarfrompsd}. For all resonator tracking schemes, $\cS_{y_{\textsf{\tiny r}}}\pqty{\omega}$ has the form of low-pass filtered white noise (Lorentzian spectrum), assuming detection noise is negligible. In this case, $\text{\small \textsf{AD}}$ first increases as a function of averaging time $\tau$, reaches its peak value $\sigma_y^{\textsf{\tiny max}}$  at around the response time constant (inverse of the bandwidth of $\cS_{y_{\textsf{\tiny r}}}\pqty{\omega}$), and decreases as $\tau$ is increased further~\cite{demir2019fundamental}. For $\tau$ larger than the response time constant, $\sigma_y^2\pqty{\tau}$ is determined by $\cS_{y_{\textsf{\tiny r}}}\pqty{0}$, simply given by 
\begin{equation}
\sigma_y^2\pqty{\tau} = \frac{\cS_{y_{\textsf{\tiny r}}}\pqty{0}}{\tau} 
\label{eqn:allanvarlargetau}  
\end{equation}
The above expression is valid for averaging time $\tau$ that is {\em larger} than the time-constant of the (low-pass) noise filtering mechanisms that reduce $\cS_{y_{\textsf{\tiny r}}}\pqty{\omega}$ as compared with $\cS_{y_{\textsf{\tiny r}}}\pqty{0}$ for $\omega$ larger than the filtering bandwidth.  Since this time-constant (inverse of filtering bandwidth) also determines the response time of the resonant sensor to resonance frequency jump events, the relevant time scales of practical interest are in fact represented by \eqref{eqn:allanvarlargetau}.
This justifies our use of $\cS_{y_{\textsf{\tiny r}}}\pqty{0}$ in Section~\ref{sec:accuracy} as an ultimate measure of accuracy that is independent of speed. If $\tau$ is increased even further, $\sigma_y\pqty{\tau}$ starts increasing again due to thermal drift and other slow nonideal drift phenomena~\cite{sadeghi2020frequency}. The behavior of $\sigma_y\pqty{\tau}$ for the three time scales is illustrated in Fig.~\ref{fig:fomstr} (upper right graph). 
For these illustrations, we assumed that the detection noise is negligible, and thermomechanical noise is dominant.

The case when detection noise is not negligible (while thermomechanical noise is still well resolved above the detection noise floor) and the detection bandwidth (of the filter $H_\textsf{\tiny L}\pqty{\las}$) is larger than the resonator linewidth was experimentally investigated in~\cite{sadeghi2020frequency}. In this scenario, $\cS_{y_{\textsf{\tiny r}}}\pqty{\omega}$, the spectrum of fractional frequency fluctuations in \eqref{eqn:specy}, exhibits low-pass filtered white noise behavior at two different time scales for the {\sc FF} scheme. That is, $\cS_{y_{\textsf{\tiny r}}}\pqty{\omega}$ is the sum of two Lorentzian spectra, one for thermomechanical noise with a corner frequency equal to the resonator linewidth, and the other for detection noise with a corner frequency equal to the detection bandwidth. In this case,  $\sigma_y\pqty{\tau}$ exhibits two peaks and a valley between them, followed by an eventual increase due to drift. The first peak (at smaller $\tau$) occurs at the detection filter time constant, whereas the second peak occurs at the resonator response time. Behavior in $\sigma_y\pqty{\tau}$ as such for the {\sc FF} scheme predicted by our theory was shown in~\cite{sadeghi2020frequency} to agree very well with experimental results. The depth of the valley between the two peaks is determined by the detection noise level with respect to thermomechanical noise~\cite{sadeghi2020frequency}.    

The problem with $\text{\small \textsf{AD}}$ is that it misleadingly suggests better accuracy performance for $\tau$ smaller than the response time of the resonant sensor. In fact, it is not meaningful to use $\text{\small \textsf{AD}}$ in this region for accuracy characterization. $\text{\small \textsf{AD}}$ is appropriate for characterizing random fluctuations in frequency, as well as deterministic frequency drifts, but not for deterministic frequency {\em offsets}. In fact, $\text{\small \textsf{AD}}$ simply evaluates to zero when there is a deterministic frequency offset, \ie a bias error.  Thus, the bias error needs to be characterized separately. This can be readily done via {\sc FSTR}  $f_{\textsf{\tiny str}}\pqty{\tau}$, defined by \eqref{eqn:fstr}, normalized
so that $f_{\textsf{\tiny str}}\pqty{0}=0$ and $f_{\textsf{\tiny str}}\pqty{\tau\rightarrow\infty} = 1$. The frequency estimate provided by any resonant sensor scheme has a bias 
\begin{equation}
\text{\bf bias}\pqty{\tau} =  \Delta y\,\bqty{1-f_{\textsf{\tiny str}}\pqty{\tau}}
\end{equation} 
in addition to the random fluctuations characterized by $\text{\small \textsf{AD}}$, where 
$\Delta y$ is the root cause amount of fractional frequency shift.  
{\em Root Mean Squared Error} ($\text{\small \textsf{RMSE}}$) which combines the two is defined as follows 
\begin{align}
\textsf{\small RMSE}\pqty{\tau} =  \sqrt{\sigma_y^2\pqty{\tau} + {{\text{\bf bias}}^{2}\pqty{\tau}}}
\label{eqn:mse}
\end{align}   
For $\tau$ less then the response time, $\textsf{\small RMSE}$ is dominated by the bias term. As $\tau$ is increased, $f_{\textsf{\tiny str}}\pqty{\tau}$ increases, and hence $\bqty{1-f_{\textsf{\tiny str}}\pqty{\tau}}$ decreases. The bias becomes small and then negligible when $\tau$ exceeds the response time, where \textsf{\small RMSE} becomes equal to $\sigma_y\pqty{\tau}$. 

Fig.~\ref{fig:fomstr} graphically illustrates the behaviors of {\sc FSTR}, $\text{\small \textsf{AD}}$ and \textsf{\small RMSE}, for the {\sc FF}, {\sc FLL} and {\sc SSO} schemes. We recall that, for these illustrations, we assumed that the impact of detection noise is negligible as compared with the thermomechanical noise of the resonator, and hence the thermomechanical noise is well resolved above the detection noise floor. {\sc FLL} and {\sc SSO} have the same performance as derived before. Fig.~\ref{fig:fomstr} illustrates and compares the performance  of {\sc FF} (with bandwidth fixed at the resonator linewidth) with {\sc FLL} and {\sc SSO} as the bandwidth for these schemes is increased. As seen in these illustrations, using $\text{\small \textsf{AD}}$ alone for performance characterization misleadingly suggests that the {\sc FF} scheme has better performance for small values of $\tau$. However, \textsf{\small RMSE} correctly indicates better performance for {\sc FLL} and {\sc SSO} schemes for time scales below the response time of  {\sc FF}. On the other hand, all schemes have the same performance at long time scales beyond the response time of {\sc FF}.  
 
We note that another figure-of-merit called  $\text{\small \textsf{FSTDEV}}$ (Frequency STep DEViation) that also combines {\sc FSTR} with $\text{\small \textsf{AD}}$ was proposed in~\cite{demir2019fundamental}. Upon further and careful evaluation, we concluded that  $\text{\small \textsf{FSTDEV}}$ has some deficiencies in appropriately characterizing performance for resonant sensor schemes with varying bandwidths. On the other hand, $\text{\small \textsf{RMSE}}$ is a universal, well established error measure for estimators that takes into account both random deviations due to noise and biases, \ie deterministic offsets.

\section{Conclusions}
\label{sec:conc}
We have presented an in-depth theory and comparative analyses for the main resonant sensor schemes that are currently in use, and unraveled their speed versus accuracy trade-off characteristics. We have developed accurate but simple and analytically tractable models, and attained a deep understanding of various issues in designing and optimizing resonant sensors. The predictions based on the theory presented in this paper for various scenarios as system parameters are varied, for both the {\sc FF} and {\sc FLL} schemes, were shown to agree very well with experimental results~\cite{sadeghi2020frequency}. Even though we have focused on nanomechanical resonant sensors in this paper, our treatment and the proposed techniques are generally applicable to all kinds of resonant sensors in other domains, \eg optics or microwaves, where the sensing scheme is based on tracking the resonance frequency of a resonator.      

\section{Acknowledgments}
The author would like to thank Selim Hanay, Matt Matheny, Selim Olcum, Pedram Sadeghi, Silvan Schmid, Guillermo Villanueva and the attendees of NMC 2019 in Lausanne for many helpful discussions on nanomechanical resonant sensors.   

\bibliographystyle{IEEEtran}
\bibliography{ref}

\begin{IEEEbiography}[{\includegraphics[width=1.0in,height=1.25in,clip,keepaspectratio]{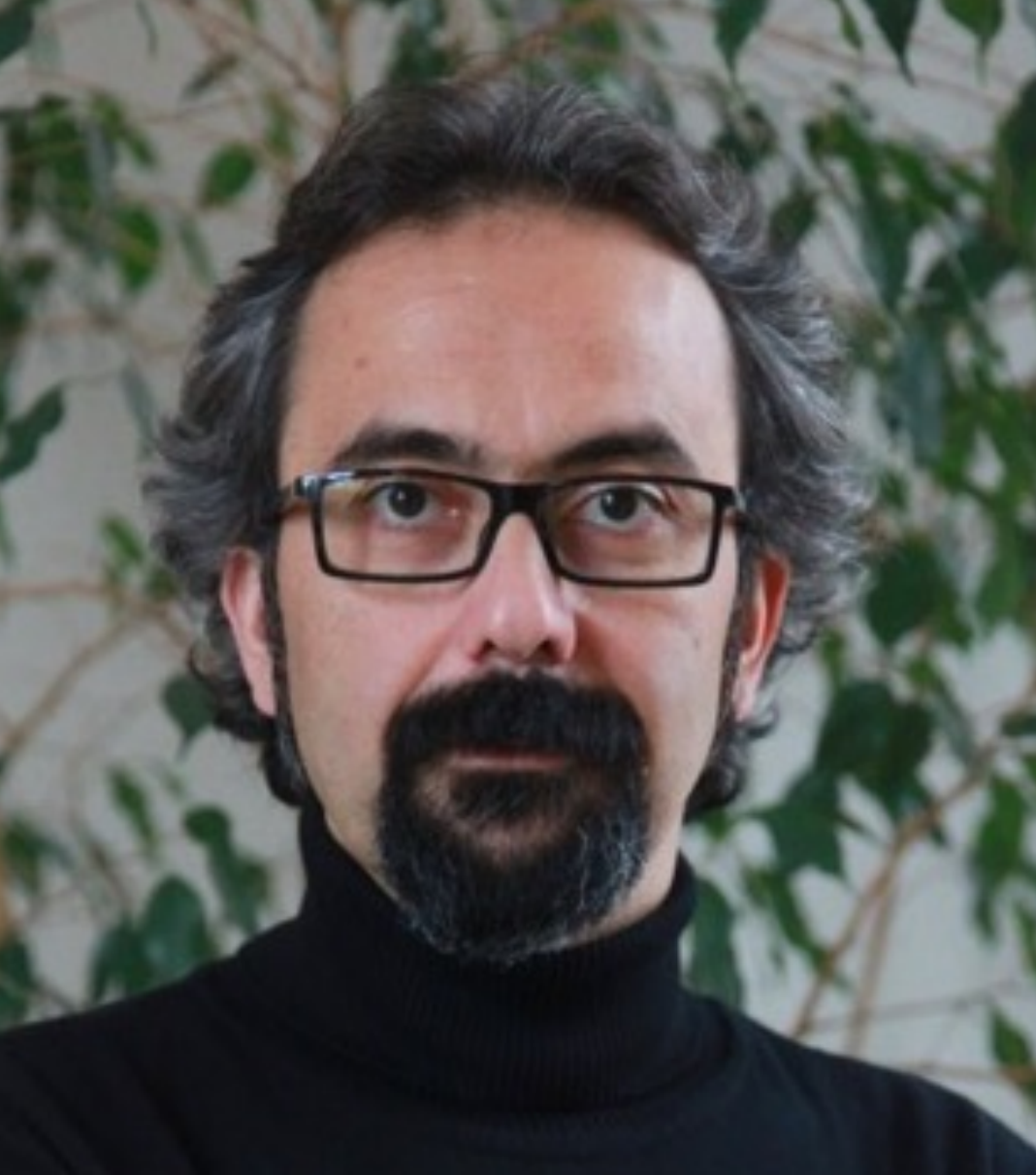}}]
{Alper Demir}
received the B.S. degree from Bilkent University, Ankara,
Turkey, in 1991, and the M.S. and Ph.D. degrees from the University of
California, Berkeley, CA, USA, in 1994 and 1997.  He was
with Motorola, Austin, TX, USA, Cadence Design Systems, San Jose, CA,
USA, Bell Laboratories, Murray Hill, NJ, USA, MIT, Cambridge, MA, USA,
and University of California at Berkeley. He has been a Faculty Member
with Ko\c{c} University, Istanbul, Turkey, since 2002.  Dr. Demir was a
recipient of the 2002 Best of ICCAD Award, the 2003/2014 IEEE/ACM
William J. McCalla ICCAD best paper awards, and the 2004 IEEE
Guillemin-Cauer Award. He was named an IEEE Fellow in 2012. 
\end{IEEEbiography}

\end{document}